\documentclass{article}[12pt]
\usepackage{amsmath,amssymb}
\usepackage{graphicx}%
\usepackage{xcolor}
\usepackage{verbatim}
\usepackage{multirow}

\newtheorem{stat}{Statment}

\newcommand{\MI}{$\mathfrak{M}_1$}

\newcommand{\dt}[1]{\frac{d#1}{dt}}

\newcommand{\fig}[4]{%
\begin{center}
\parbox{#2cm}{%
\refstepcounter{figure}\includegraphics[width=#2cm,height=#3cm]{#1}\\ \noindent Fig. \thefigure:\quad
#4}\end{center}}
\newcommand{\TextFigReg}[5]{%
\begin{flushleft}
\begin{tabular}{lcr}
\parbox{#4cm}{#5} & \hskip 10mm &
\parbox{#2cm}{\includegraphics[width=#2cm]{#1}\\[12pt]
\refstepcounter{figure}Fig. \thefigure.\quad #3\vfill}  \\
\end{tabular}
\end{flushleft}
\vspace{7pt}
}

\newcommand{\TwoFigs}[4]{%
\begin{flushleft}
\begin{tabular}{cc}
\parbox{7cm}{\centerline{\includegraphics[width=7cm]{#1}}}  & \parbox{7cm}{\centerline{\includegraphics[width=7cm]{#2}}}  \\
\parbox{7cm}{\vspace{7pt}\refstepcounter{figure}Fig. \thefigure.\quad #3\vfill} & \parbox{7cm}{\vspace{7pt}\refstepcounter{figure}Fig. \thefigure.\quad #4\vfill}\\
\end{tabular}
\end{flushleft}
\vspace{7pt}
}

\newcounter{strochka}
\newcommand{\stroka}[1]{\refstepcounter{strochka}\par\noindent\textsl{\arabic{strochka}}. \; \textsl{#1}}
\newcounter{spisok}
\begin{document}

\begin{center}
{\bf \Large Yu.G. Ignat'ev\footnote{Institute of Physics, Kazan Federal University, Kremlyovskaya str., 35, Kazan, 420008, Russia; email: yurii.ignatev.1947@yandex.ru} }\\[12pt]
{\bf \Large Evolution of spherical perturbations in the cosmological environment of degenerate scalarly charged fermions with the Higgs scalar interaction} \\[12pt]
\end{center}

\abstract{A mathematical model is constructed for the evolution of spherical perturbations in a cosmological one-component statistical system of completely degenerate scalarly charged fermions with a scalar Higgs interaction. A complete system of self-consistent equations for small perturbations describing the evolution of spherical perturbations is constructed. Singular parts in perturbation modes corresponding to point mass and scalar charge are singled out. Systems of ordinary differential equations are obtained that describe the evolution of the mass and charge of a singular source, and systems of partial differential equations that describe the evolution of non-singular parts of perturbations. In this case, the coefficients of partial differential equations are described by solutions of evolutionary equations for mass and charge. The problem of spatially localized perturbations for solutions polynomial in the radial coordinate is reduced to a recurrent system of ordinary linear differential equations for the coefficients of these polynomials. The properties of solutions in the case of cubic polynomials are studied, in particular, it is shown that the radii of localization of gravitational and scalar perturbations coincide and evolve in proportion to the scale factor. Numerical modeling of the evolution of perturbations was carried out, in particular, it confirmed the exponential growth of the central mass of the perturbation, and also revealed the oscillatory nature of the evolution of the scalar charge.

\textbf{Keywords}: scalar charged plasma, cosmological model, scalar Higgs field, gravitational instability, spherical perturbations.
}


\section*{Introduction}
In the papers \cite{Ign_GC21_Un} -- \cite{Yu_GC_23} we considered in detail the problem of the formation of nuclei of supermassive black holes (BHS) with a mass of the order of \cite{Zhu}
\begin{equation}\label{M_nc}
M_{bhs}\sim 10^4\div 10^6 M_\odot\approx 10^{42}\div10^{44}m_{\mathrm{pl}}.
\end{equation}
at redshifts $z\gtrsim7$, which exists in astrophysics and cosmology. The papers \cite{SMBH1} -- \cite{Soliton}, which are key for this problem, are considered in the same articles. Therefore, in this paper, we will not return to the issues of substantiating the ongoing research, referring the Reader to the above articles.

In the work cited above \cite{Ign_GC21_Un} based on the complete theory of a two-component system of degenerate scalarly charged fermions\footnote{Such fermions, like other particles, can appear at the inflationary stage of the expansion of the Universe as a result of the processes of particle production by gravitational and scalar fields.} with Higgs scalar fields \cite{TMF_21} and the results of numerical modeling of the corresponding cosmological model, a numerical model of the evolution of scalar-gravitational perturbations was constructed for the case of an asymmetric scalar doublet \MI, examples of the development of instability in the cosmological system were given, and some features of this process were revealed. Further, in the papers \cite{Yu_GC_3_22} -- \cite{YU_GC_4_22} a systematic study of the development of scalar-gravitational perturbations in a cosmological model based on a single-component system of degenerate scalar-charged fermions with the classical Higgs interaction was carried out with a view to the possibility of the formation of supermassive black holes in early universe. In this case, the processes of their Hogging evaporation were taken into account. Studies have confirmed the fundamental possibility of the early formation of black holes with the required masses \eqref{M_nc}. Further, to clarify the connection between instability and bifurcation points of the vacuum-scalar cosmological model, \cite{Yu_GC_23} studied the influence of a phantom field on the development of gravitational instability. It is shown that the scalar-gravitational instability in the early stages of expansion in the model under study arises at sufficiently large scalar charges, and the instability develops precisely near the unstable points of the vacuum doublet. In this case, short-wave perturbations even of a free phantom field turn out to be stable at stable singular points of the vacuum doublet. The key point of the theory of scalar-gravitational instability proposed in \cite{Yu_GC_3_22} -- \cite{Yu_GC_23} is the exponentially fast growth of linear perturbations with time, which, unlike the standard Lifshitz power law for gas-liquid models, allows perturbations to grow to sufficiently large values over a very short interval of cosmological time. The reason for the rapid development of perturbations is, apparently, a combination of two factors - the interparticle scalar attraction of charged fermions when critical densities are reached, and the macroscopic gravitational attraction.

Although these works substantiate the fundamental possibility of the early formation of supermassive black holes, however, the corresponding hypothesis requires additional substantiation, in particular, studies of the evolution of spherical perturbations, which should correspond to the processes of formation of black holes. This article is devoted to the development of a theoretical and mathematical model of this process and methods for its study. In this case, the key role will be played, firstly, by the \emph{method of invariant selection of the singular part of perturbations}, developed in the papers \cite{YuPhys_83} -- \cite{YuPhys3_08} and allowing one to separate the variables, split the system of differential equations into perturbations, and then reduce it to two subsystems - an autonomous system of ordinary differential equations for the singular part of perturbations and an independent system of relatively nonsingular perturbations. Secondly, the key role in the developed mathematical model will be played by the concept of a spatially localized spherical perturbation introduced in the same works, thanks to which the problem can be reduced to a closed chain of ordinary linear differential equations.

In this case, in contrast to previous works, we will not restrict ourselves to the short-wavelength approximation, without imposing any conditions on the perturbation lengths.

\section{Mathematical model\newline of degenerate fermions with scalar interaction}

In \cite{TMF_21} shows how, based on the Lagrangian formalism, from the microscopic equations of motion of scalarly charged particles, one can obtain a macroscopic model of a statistical system of scalarly charged particles, described by macroscopic flows. In this article, we will use the results obtained in \cite{TMF_21}.

Below we will consider a cosmological model based on a one-component degenerate statistical system of scalarly charged fermions and a scalar Higgs field $\Phi$.
The dynamical mass $m_z$ of fermions $z$ with scalar charge $e$ with respect to the canonical field $\Phi$ is described by the formula:
\begin{equation}\label{mF}
m_z=e\Phi.
\end{equation}
The Lagrange function $L_s$ of the scalar Higgs field is\footnote{Hereinafter, Latin letters run over $\overline{1,4}$, Greek ones $\overline{1,3}$. The Planck system of units $G=c=\hbar=1$ is used throughout.
\label{Plank_units}}
\begin{eqnarray} \label{L_s}
L_s=\frac{1}{16\pi}(g^{ik} \Phi_{,i} \Phi_{,k} -2V(\Phi)),
\end{eqnarray}
where
\begin{eqnarray}
\label{Higgs}
V(\Phi)=-\frac{\alpha}{4} \left(\Phi^{2} -\frac{m_0^{2}}{\alpha}\right)^{2}
\end{eqnarray}
is the potential energy of the scalar field, $\alpha$ is the self-action constant, $m_0$ is the mass of quanta.
The energy-momentum tensor of scalar fields with respect to the Lagrange function \eqref{L_s} is:
\begin{eqnarray}\label{T_s}
S^i_{k} =\frac{1}{16\pi}\bigl(2\Phi^{,i}\Phi_{,k}- \delta^i_k\Phi_{,j} \Phi^{,j}+2V(\Phi)\delta^i_k \bigr),
\end{eqnarray}
Further, the energy-momentum tensor \emph{equilibrium} of the statistical system is equal to:
\begin{equation}\label{T_p}
T^i_{k}=(\varepsilon_p+p_p)u^i u_k-\delta^i_k p_p,
\end{equation}
where $u^i$ is the macroscopic velocity vector of the statistical system.

Einstein's equations for the ``scalar field+particles'' system are:
\begin{equation}\label{Eq_Einst_G}
R^i_k-\frac{1}{2}\delta^i_k R=8\pi (T^i_k+S^i_k) + \delta^i_k \Lambda_0,
\end{equation}
where $\Lambda_0$ is the bare value of the cosmological constant, related to its observed value $\Lambda$ obtained by removing the constant terms in the potential energy by the relation:
\begin{equation}\label{lambda0->Lambda}
\Lambda=\Lambda_0-\frac{1}{4}\frac{m_0^4}{\alpha}.
\end{equation}

Strict macroscopic consequences of the kinetic theory are the transport equations, including the conservation law of some vector current corresponding to the microscopic conservation law in the reactions of some fundamental charge ${\rm Q}$ with particle charges $q$ -- %
\begin{equation}\label{1}
\nabla_i q n^i=0,
\end{equation}
as well as the laws of conservation of energy - momentum of the statistical system:
\begin{equation}\label{2}
\nabla _{k} T_{p}^{ik} -\sigma\nabla^{i} \Phi =0,
\end{equation}
where $\sigma$ is the density of scalar charges with respect to the field $\Phi$ \cite{TMF_21}.

The well-known identity follows from the velocity vector normalization relation
\begin{equation}\label{6}
u^k_{~,i}u_k\equiv 0,
\end{equation}
which allows us to reduce the laws of conservation of energy - momentum (\ref{2}) to the form of equations of ideal hydrodynamics
\begin{eqnarray}\label{2a}
(\varepsilon_p+p_p)u^i_{~,k}u^k=(g^{ik}-u^iu^k)(p_{p,k}+\sigma\Phi_{,k});\\
\label{2b}
\nabla_k[(\varepsilon_p+p_p)u^k]=u^k(p_{p,k}+\sigma\Phi_{,k}),
\end{eqnarray}
and the laws of conservation of the fundamental charge $\mathrm{Q}$ \eqref{1} to the form
\begin{equation}\label{2c}
\nabla_k \rho u^k=0,
\end{equation}
where $\rho\equiv q n$ -- \emph{kinematic density of the scalar charge} of the statistical system with respect to the scalar field $\Phi$.

Macroscopic scalars for a one-component statistical system of degenerate fermions take the form:
\begin{eqnarray}
\label{2_3c}
n=\frac{1}{\pi^2}\pi_z^3;\; p_p  =\displaystyle \frac{e^4\Phi^4}{24\pi^2}(F_2(\psi)-4F_1(\psi));\\
\label{2_3a_2}
 \sigma=\frac{e^4 \Phi^3}{2\pi^2}F_1(\psi);   \; \varepsilon_p=\frac{e^4 \Phi^4}{8\pi^2}F_2(\psi),
\end{eqnarray}
where $\pi_z$ is the Fermi momentum, $\sigma$ is the density of scalar charges $e$ and
\begin{equation}\label{psi}
\psi=\frac{\pi_z}{|e\Phi|}.
\end{equation}
Functions $F_1(\psi)$ and $F_2(\psi)$ are introduced to shorten the letter:
\begin{eqnarray}\label{F_1}
F_1(\psi)=\psi\sqrt{1+\psi^2}-\ln(\psi+\sqrt{1+\psi^2});\nonumber\\
\label{F_2}
\!\!\!F_2(\psi)=\psi\sqrt{1+\psi^2}(1+2\psi^2)-\ln(\psi+\sqrt{1+\psi^2}).\nonumber
\end{eqnarray}
Thus, we have the relation
\begin{equation}\label{{e+p}_p}
(\varepsilon+p)_p=\frac{e^4\Phi^4}{3\pi^2}\psi^3\sqrt{1+\psi^2}.
\end{equation}
Finally, the scalar field equations for a one-component system take the form:
\begin{eqnarray}\label{Box(Phi)=sigma_z}
\Box \Phi + m_0^2\Phi-\alpha\Phi^3 =-8\pi\sigma\equiv-\frac{4e^4\Phi^3}{\pi} F_1(\psi).
\end{eqnarray}

\subsection{Model transformation properties}
Let us consider the complete system of equations of the mathematical model, which consists of the Einstein equations \eqref{Eq_Einst_G}, the equations of hydrodynamics \eqref{2} and the equation of the scalar field \eqref{Box(Phi)=sigma_z} together with the definitions of the corresponding sources: the energy-momentum tensor of the scalar field , \eqref{T_s}, fermionic liquid, \eqref{T_p} and scalar charge density, \eqref{2_3a_2}, as well as definitions of fermion energy density \eqref{2_3c} and their pressure \eqref{2_3a_2}. As can be seen from the equations of this system and the definitions of its coefficients, the solutions of the Cauchy problem for this system of equations are completely determined by the corresponding initial conditions for the metric functions1 $g_{ik}(x^j)$, the potential $\Phi(x^j)$, the vector velocity $u^i(x^j)$ and Fermi momentum $\pi_z$.

Consider the following \emph{scaling} transformations of \emph{fundamental model parameters} $P=[\alpha,m_0,e,\Lambda_0]$
\begin{eqnarray}\label{trans_param}
\alpha\to k^2\alpha,\; m_0\to km_0;\; e\to \sqrt{k}e;\;\Lambda_0\to k^2\Lambda_0,\quad (k=\mathrm{Const}>0).
\end{eqnarray}
Along with the transformations of the fundamental parameters, we will perform scaling transformations of the coordinates of the mathematical model
\begin{eqnarray}\label{x_trans}
x^i\to k^{-1}x^i,\quad \pi_z\to \sqrt{k}\pi_z, \Phi\to\Phi, \; u^i\to u^i,
\end{eqnarray}
as well as the Fermi momentum and the scalar potential
\begin{equation}\label{(pi,Phi)_trans}
\pi_z\to \sqrt{k}\pi_z;\quad \Phi\to\Phi.
\end{equation}

Obviously, under scaling transformations \eqref{trans_param} -- \eqref{(pi,Phi)_trans} the scalars and tensors introduced above change according to the laws:
\begin{eqnarray}
\psi\to\psi,\; \sigma\to k^2\sigma;\;V(\Phi)\to k^2 V(\Phi);\; p_p\to k^2 p_p;\; \varepsilon_p\to k^2 \varepsilon_p;\nonumber\\
S^i_k\to k^2 S^i_k;\quad T^i_k\to k^2 T^i_k.
\end{eqnarray}
Thus, the laws of scaling transformation of the equations of the mathematical model are valid
\begin{eqnarray}\label{Eq_Einst_trans}
R^i_k-\frac{1}{2}\delta^i_k R=8\pi (T^i_k+S^i_k) + \delta^i_k \Lambda_0\rightarrow k^2\bigl(R^i_k-\frac{1}{2}\delta^i_k R=8\pi (T^i_k+S^i_k) + \delta^i_k \Lambda_0\bigr);\\
\label{Box(Phi)=sigma_z_trans}
\Box \Phi + m_0^2\Phi-\alpha\Phi^3 =-8\pi\sigma\rightarrow k^2\bigl(\Box \Phi + m_0^2\Phi-\alpha\Phi^3 =-8\pi\sigma\bigr);\\
\label{2_trans}
\nabla _{k} T_{p}^{ik} -\sigma\nabla^{i} \Phi =0\rightarrow k^2\bigl(\nabla _{k} T_{p}^{ik} -\sigma\nabla^{i} \Phi =0\bigr).
\end{eqnarray}

Thus, the following \emph{similarity property of a mathematical model} is true.
\begin{stat}\label{stat1}
The complete system of equations of the mathematical model \eqref{Eq_Einst_G}, \eqref{2}\emph{ and} \eqref{Box(Phi)=sigma_z} is invariant under scaling transformations of the fundamental parameters of the mathematical model \eqref{trans_param} and scaling coordinate transformations \eqref{x_trans} \emph{ and the Fermi momentum }\eqref{(pi,Phi)_trans}, i.e., the solutions of the equations of the original model and the scale-transformed one coincide.
\begin{eqnarray}\label{trans_eqs}
\Phi\to\Phi;\qquad g_{ik}\to g_{ik};\quad u^i\to u^i.
\end{eqnarray}
\end{stat}
This important similarity property of a mathematical model allows one to extend the solution with a given set of fundamental parameters to the case of other values of fundamental parameters.

\section{Linear spherical perturbations of the cosmological model}
\cite{Ignat_STFI}, \cite{Ignat_Phys_22_1}, \cite{Ignat_Phys_22_2} developed a general theory of the evolution of linear \emph{planar} scalar-gravitational perturbations in a two-component cosmological system of degenerate scalar-charged fermions with asymmetric Higgs scalar interaction. In this article, we consider the theory of evolution of linear spherical perturbations in a one-component cosmological system of degenerate scalarly charged fermions with a canonical Higgs scalar interaction.

\subsection {Unperturbed isotropic homogeneous ground state}
As a background, we consider the spatially flat Friedman metric
\begin{eqnarray}\label{ds_0}
ds_0^2=dt^2-a^2(t)(dx^2+dy^2+dz^2)\equiv \\
dt^2-a^2(t)[dr^2+r^2(d\theta^2+\sin^2\theta d\varphi^2)],\nonumber
\end{eqnarray}
and as a background solution we consider a homogeneous isotropic distribution of matter, in which all thermodynamic functions and scalar fields depend only on the cosmological time $t$:
\begin{equation}\label{base_state}
\Phi=\Phi(t);\;  \pi_z=\pi_z(t);\;  u^i=u^i(t).
\end{equation}
Note that the physically measured radius in the \eqref{ds_0} metric is
\begin{equation}\label{R=}
R=a(t)r.
\end{equation}
It is easy to verify that
\begin{equation}\label{u_0}
u^i=\delta^i_4
\end{equation}
converts the equations (\ref{2a}) into identities, and the system of equations (\ref{2b}) -- (\ref{2c}) reduces to 2 \emph{material} equations:\footnote{Here and
below $ \dot{f}=\partial f/\partial t$, $f'=\partial f/\partial r$.}
\begin{equation}\label{7a1-0}
\dt{\varepsilon_p}+3\frac{\dot{a}}{a}(\varepsilon_p+p_p)=\sigma\dot{\Phi};
\end{equation}
\begin{eqnarray}\label{7b1-0}
\dt{n} +3\frac{\dot{a}}{a}n=0.
\end{eqnarray}
In \cite{TMF_21} shows that the \eqref{7a1-0} -- \eqref{7b1-0} system of equations has simple solutions:
\begin{equation}\label{aP0}
 a\pi_z=\mathrm{const}.
\end{equation}
Taking \eqref{aP0} into account, we write the dimensionless function $\psi$ \eqref{psi} explicitly:
\begin{equation}\label{psi(eta)}
\psi=\frac{\pi^0_z}{|e\Phi|}\mathrm{e}^{-\xi}, \quad (\pi^0_z=\pi_z(0)),
\end{equation}
where we have passed to the new variable $\xi(t)$
\begin{equation}\label{a-xi}
\xi=\ln a,
\end{equation}
assuming here and in the future
\begin{equation}\label{xi(0)}
\xi(0)=0.
\end{equation}

Further, the energy-momentum tensor of the scalar field in the unperturbed state also takes the form of the energy-momentum tensor of an ideal isotropic fluid:
\begin{equation} \label{MET_s}
S^{ik} =(\varepsilon_s +p_{s} )u^{i} u^{k} -p_s g^{ik} ,
\end{equation}
wherein:
\begin{eqnarray}\label{Es}
\varepsilon_s=\frac{1}{8\pi}\biggl(\frac{1}{2}\dot{\Phi}^2+V(\Phi)\biggr);\\
\label{Ps}
p_{s}=\frac{1}{8\pi}\biggl(\frac{1}{2}\dot{\Phi}^2-V(\Phi)\biggr),
\end{eqnarray}
so:
\begin{equation}\label{e+p}
\varepsilon_s+p_{s}=\frac{1}{8\pi}\dot{\Phi}^2.
\end{equation}
The equation of the unperturbed scalar field \eqref{Box(Phi)=sigma_z} in the Friedmann metric takes the form:
\begin{eqnarray}\label{Eq_Phi_eta}
\ddot{\Phi}+\frac{3}{a}\dot{a}\dot{\Phi}+m_0^2\Phi-\alpha\Phi^3= -8\pi\sigma,
\end{eqnarray}
where the scalar charge density $\sigma$ is described by the expression \eqref{2_3c}, in which it is necessary to substitute the value of the function
$\psi$ \eqref{psi(eta)}.
As a result, the complete autonomous system of equations of the unperturbed cosmological model takes the form \cite{TMF_21}:
\begin{eqnarray}\label{dot_xi-dot_varphi}
\dot{\xi}=H; \; \dot{\Phi}=Z;\;\\
\label{dH/dt_0}
\dot{H}=- \frac{Z^2}{2}-\frac{4}{3\pi}e_z^4\Phi^4\psi^3\sqrt{1+\psi^2};\\
\label{dZ/dt}
\dot{Z}=-3HZ-m_0^2\Phi +\Phi^3\biggl(\alpha-\frac{4e^4}{\pi}F_1(\psi)\biggr),
\end{eqnarray}
where $H(t)$ is the Hubble parameter
\begin{equation}\label{H}
H= \frac{\dot{a}}{a}\equiv \dot{\xi}.
\end{equation}
Moreover, the Einstein equation $^4_4$ is the first integral of the system \eqref{dot_xi-dot_varphi} -- \eqref{dZ/dt}:
\begin{eqnarray}\label{Surf_Einst1_0}
3H^2-\Lambda-\frac{Z^2}{2}-\frac{m_0^2\Phi^2}{2}+\frac{\alpha\Phi^4}{4}-\frac{e^4\Phi^4}{\pi}F_2(\psi)=0.
\end{eqnarray}

In \cite{Yu_GC_3_22}, \cite{Ignat_GC21}, \cite{Ignat_GC21a}, as well as in the earlier work \cite{Yu_Sasha_Dima_ASS}, a collection of behaviors of cosmological models based on the system of equations \eqref{dot_xi-dot_varphi} - - \eqref{Surf_Einst1_0} for different values of fundamental parameters and initial conditions, as well as models based on a two-component system of fermions. Among these models there are models with a finite evolution time, passing from the expansion stage to the contraction stage, as well as models that maintain the value of the Hubble parameter close to zero for a significant time at intermediate stages of cosmological evolution.
\subsection{Expansion in perturbations}
We write the metric with gravitational perturbations in isotropic spherical coordinates with the conformally Euclidean metric of three-dimensional space\footnote{see, for example, \cite{Land_Field}}, which admits a continuous transition to the Friedmann metric \eqref{ds_0} \cite{YuPhys1_08}:
\begin{eqnarray}
\label{metric_pert}
ds^2=\mathrm{e}^{\nu(r,t)}dt^2-a^2(t)\mathrm{e}^{\lambda(r,t)}[dr^2+r^2(d\theta^2+\sin^2\theta d\varphi^2)],
\end{eqnarray}
where $\nu(r,t)$ and $\lambda(r,t)$ are small longitudinal perturbations of the Friedmann metric ($\nu\ll1,\ \lambda\ll1$).

Next, we introduce perturbations of fermions and a scalar field according to \cite{GC_21_1}
\begin{eqnarray}\label{dF-drho-du}
\begin{array}{lcl}
\Phi(r,t)&=&\Phi(t)+\varphi(r,t);\\
\pi_z(r,t)&=&\pi_z(t)(1+\delta_z(r,t));\\
\sigma(r,t)&=& \sigma(t)+\delta\sigma(r,t);\\
u^i&=&\displaystyle \delta^i_4(1-\nu(r,t))+\delta^i_1 v(r,t),\\
\end{array}
\end{eqnarray}
where $\varphi(r,t)$, $\delta_z(r,t)$, $\delta\sigma(r,t)$ and $v(r,t)$ are functions of the first order of smallness compared to their unperturbed values.
%
%

%
Computing according to \eqref{2_3c} -- \eqref{2_3a_2} perturbations of macroscopic scalars, taking into account the useful relation
\begin{equation}\label{dpsi}
\delta\psi=\psi\left(\delta_z(r,t)-\frac{\varphi(r,t)}{\Phi}\right),
\end{equation}
we find:
\begin{eqnarray}\label{dn}
\delta n= \displaystyle 3n(\eta) \delta_z(r,t);\\
\label{dsigma}
\delta\sigma= \frac{e^4\Phi^3}{2\pi^2}\biggl[\frac{\psi^3}{\sqrt{1+\psi^2}}\delta_z
-\biggl(3F_1(\psi)-\frac{2\psi^3}{\sqrt{1+\psi^2}}\biggr)\frac{\varphi}{\Phi}\biggr];\\
\label{de}
\delta\varepsilon_p= \frac{e^4\Phi^4}{\pi^2}\biggl[\psi^3\sqrt{1+\psi^2}\delta_z+
\biggl(\frac{1}{2}F_2(\psi)-\psi^3\sqrt{1+\psi^2}\biggr)\frac{\varphi}{\Phi}\biggr];\\
\label{dp}
\delta p_p= \frac{e^4\Phi^4}{3\pi^2}\biggl[\frac{\psi^5}{\sqrt{1+\psi^2}}\delta_z +
 \biggl(\frac{1}{2}(F_2(\psi)-4F_1(\psi))-\frac{\psi^5}{\sqrt{1+\psi^2}}\biggr)\frac{\varphi}{\Phi}\biggr].
\end{eqnarray}

\subsection{Perturbations of the energy-momentum tensor}
According to \eqref{T_p} and \eqref{dF-drho-du}, in the linear approximation, fermion EMT perturbations have the following nonzero components:
\begin{eqnarray}\label{dT}
\delta T^\alpha_{\beta(p)}=-\delta^\alpha_\beta\delta p_p;\;
\delta T^4_{4(p)}=\delta\varepsilon_p; \\
\delta T^\alpha_{4(p)}=-\frac{1}{a^2}\delta T^4_{\alpha(p)}=\upsilon(\varepsilon+p)_p \delta^\alpha_1.\nonumber
\end{eqnarray}

Taking \eqref{MET_s}, \eqref{Es}, \eqref{Ps} and \eqref{dF-drho-du} into account, we obtain for the non-zero components of perturbations of the scalar field energy-momentum tensor \eqref{T_s}
\begin{eqnarray}\label{dS}
\delta S^\alpha_\beta=\frac{\delta^\alpha_\beta}{8\pi}\biggl[\frac{\nu}{2}Z^2-\dot{\varphi}Z+\varphi(m_0^2-\alpha\Phi^2)\Phi \biggr];\nonumber\\
\delta S^4_4=\frac{1}{8\pi}\biggl[-\frac{\nu}{2}Z^2+\dot{\varphi}Z+\varphi(m_0^2-\alpha\Phi^2)\Phi \biggr];\nonumber\\
\delta S^1_4=-\frac{\partial}{\partial r}\biggl(\frac{Za\varphi}{8\pi a^3}\biggr).
\end{eqnarray}

Further, in order for the perturbed solution at a distance to smoothly pass into the background cosmological solution \eqref{ds_0}, \eqref{base_state}, it is necessary that all perturbations of zero conditions at infinity be satisfied
\begin{eqnarray}\label{delta(8)=0}
\lambda(\infty,t)=\nu(\infty,t)=\varphi(\infty,t)=\delta_z(\infty,t)=v(\infty,t)=0;
\end{eqnarray}
\begin{eqnarray}
\label{delta'(8)}
\lambda'(\infty,t)=\nu'(\infty,t)=\varphi'(\infty,t)=\delta'_z(\infty,t)=v'(\infty,t)=0.
\end{eqnarray}
Taking into account the symmetry of the perturbations of the fermion momentum energy tensor \eqref{dT} and the scalar field \eqref{dS}, a consequence of the Einstein equations with respect to the metric \eqref{metric_pert} is the relation (see, for example, \cite{YuPhys1_08})
\[\frac{\partial}{\partial r}\frac{1}{r}\frac{\partial}{\partial r}(\lambda+\mu)=0\Rightarrow \lambda+\mu=C_1(t)+C_2(t)rR^2,\]
where $C_1(t)$ and $C_2(t)$ are arbitrary functions of time. Therefore, due to the conditions \eqref{delta(8)=0} -- \eqref{delta'(8)}, it must be $C_1(t)=C_2(t)=0$.

Thus, in the linear approximation, for perturbations satisfying the zero conditions at infinity \eqref{delta(8)=0} and \eqref{delta'(8)}, the relation
\begin{eqnarray}\label{lambda+nu=0}
\lambda(r,t)+\nu(r,t)=0 \Rightarrow
\lambda(r,t)=-\nu(r,t).
\end{eqnarray}
In what follows, we will take into account the relation \eqref{lambda+nu=0}, choosing $\nu(r,t)$ as an independent function.

Further, note that the metric \eqref{metric_pert} admits an infinitesimal transformation with respect to the time variable
\begin{equation}\label{delta_t}
\nu(r,t)\to \nu(r,t)+\delta\xi(t),\; t\to t+\frac{1}{2}\int \delta\xi dt,
\end{equation}
without changing the form of this metric. This transformational symmetry of the \eqref{metric_pert} metric will be used in what follows.

In this work, we do not raise the question of the mechanism of occurrence of degenerate scalarly charged fermions, as well as their physical connection to various field-theoretic models, in particular, SU(5), therefore, we do not consider the corresponding fundamental parameters $\alpha,m_0,e$ concretize. We only assume that at the stage of cosmological evolution we are studying, such fermions are present, perhaps due to the primary production of massless particles and the subsequent production of massive particles in the reaction channels of the corresponding field theory model, for example, at the stage of quark-gluon plasma. Note that the very fact of degeneracy of fermions in the framework of this work does not play a significant role, it only helps to simplify the problem mathematically. It is important that the necessary gravitational perturbations are present at this stage. It is difficult to determine the moment of occurrence of spherical instability, since, according to the results of previous works (see, for example, \cite{YU_GC_4_22}), spherical perturbations should form at the nonlinear stage of the growth of plane perturbations as a result of the interaction of modes with different directions of the wave vector. In this case, the long-wavelength modes with the wavenumber $n\gtrsim1$ grow the fastest, and it is these modes that must survive in the competitive processes of formation of spherical perturbations.
To determine this moment of time and the perturbation parameters, it is necessary to construct a more complex, essentially nonlinear theory, at least quadratic in gravitational perturbations. The solution of this problem completely determines the solution of another problem - the final parameters of the formed BHS (see below, section \ref{r_razdel}).

\subsection{Linear perturbations of the Einstein equations}
Taking into account \eqref{metric_pert} and \eqref{lambda+nu=0}, as well as the unperturbed Einstein equations \eqref{dH/dt_0} and \eqref{Surf_Einst1_0}, we obtain for non-zero perturbation components of the Einstein tensor $G^i_k =R^i_k-1/2R\delta^i_k$
\begin{eqnarray}\label{dG}
\delta G^\alpha_\beta=-\delta^\alpha_\beta[\ddot{\nu}+4H\dot{\nu}+(\Lambda-8\pi p)\nu];\nonumber\\
\delta G^4_4=\frac{1}{a^3r^2}\frac{\partial}{\partial r}\biggl(r^2\frac{\partial a\nu}{\partial r}\biggr)-\frac{3H}{a}\frac{\partial a\nu}{\partial t};\nonumber\\
\delta G^1_4=-\frac{\partial}{\partial r}\biggl(\frac{1}{a^3}\frac{\partial a\nu}{\partial t} \biggr).
\end{eqnarray}
Thus, the nontrivial perturbed Einstein equations can be written as:

\begin{eqnarray}\label{Einst_dp}
\ddot{\nu}+4H\dot{\nu}+(\Lambda-8\pi p)\nu+\frac{\nu}{2}Z^2-\dot{\varphi}Z+\varphi(m_0^2-\alpha\Phi^2)\Phi=8\pi \delta p_p;\\
\label{Einst_de}
\frac{1}{a^3r^2}\frac{\partial}{\partial r}\biggl(r^2\frac{\partial a\nu}{\partial r}\biggr)-\frac{3H}{a}\frac{\partial a\nu}{\partial t}+\frac{\nu}{2}Z^2 -\dot{\varphi}Z-\varphi(m_0^2-\alpha\Phi^2)\Phi =8\pi\varepsilon_p;\\
\label{v=}
\upsilon=\frac{1}{8\pi(\varepsilon+p)_pa^3}\frac{\partial}{\partial r}\biggl(\frac{\partial a\nu}{\partial t}-Za\varphi\biggr),
\end{eqnarray}
-- Einstein's equation for the $^1_4$ component \eqref{v=} determines the radial velocity of fermions.

\subsection{Linear perturbation of the scalar field equation}
Calculating the linear correction to the scalar field equation \eqref{Box(Phi)=sigma_z}, taking into account \eqref{dF-drho-du} and the background field equation \eqref{dZ/dt}, we obtain the equation for the perturbation of the scalar field
\begin{eqnarray}\label{Eq_varphi}
\ddot{\varphi}+3H\dot{\varphi}-\frac{1}{a^2r^2}\frac{\partial}{\partial r}r^2\frac{\partial \varphi}{\partial r}+(m_0^2-3\alpha\Phi^2)\varphi
+2\dot{\nu}Z+8\pi\sigma\nu=-8\pi\delta\sigma.
\end{eqnarray}
\subsection{Identification of particle-like solutions}
The linearized Einstein equation \eqref{Einst_de} and the scalar field equation \eqref{Eq_varphi} include second derivatives with respect to the radius in the form of the radial part of the Laplace operator of the three-dimensional Euclidean space
\begin{equation}\label{Delta_r}
\Delta_r U\equiv \frac{1}{a^2r^2}\frac{\partial}{\partial r}\biggl(r^2\frac{\partial U}{\partial r}\biggr).
\end{equation}

Bearing in mind further consideration of particle-like solutions of equations for perturbations, we consider the canonical equations of motion of a classical gravitating point particle in a gravitational field, which corresponds to a singular point source. As a result of two competing processes - accretion of the surrounding material medium and the reverse process - the evaporation of matter, the mass of a classical point particle in the material medium cannot be constant. \cite{YuASS}, \cite{YuPhysA}, and \cite{YuPhys1_08} have developed methods for invariant separation of the singular part in the Einstein equations corresponding to a particle-like solution. Let us apply these methods to the study of our problem. The invariant Hamilton function of a classical massive particle in the form \cite{YuPhys_83}:
\begin{equation}\label{Ham}
H(x,P)=\sqrt{g^{ik}P_iP_k}-m,
\end{equation}
where $\mu(s)$ is the mass of the particle on the trajectory $\{x^i(s)$, $P_i(s)\}$. From \eqref{Ham} we obtain the generalized momentum normalization relation
\begin{equation}\label{P^2}
(P,P)=m^2(s).
\end{equation}
The relativistic canonical equations of motion of a massive particle have the form:
\begin{equation}\label{Can_Eqs}
\frac{dx^i}{ds}=\frac{\partial H}{\partial P_i};\qquad \frac{dP_i}{ds}=-\frac{\partial H}{\partial x^i}.
\end{equation}
From the first group of canonical equations we find, taking into account \eqref{P^2}
\[
\frac{dx^i}{ds}=\frac{P^i}{m}\Rightarrow g_{ik}\frac{dx^i}{ds}\frac{dx^k}{ds}=1.
\]

In a spherically symmetric metric, the solution of the \eqref{Can_Eqs} equations, which does not violate spherical symmetry, is the time line, which corresponds to the state of rest of the particle at the center of symmetry:
\begin{equation}\label{line_t}
P_r=0;\; r=0,\qquad s=t;
\end{equation}
in this case, the mass of the particle remains an arbitrary function of the coordinate time:
\begin{equation}\label{m(eta)}
m=m(t).
\end{equation}
Let us write in an invariant form the energy density $\delta\varepsilon_m$ corresponding to the singular part of the matter energy density:
\begin{equation}\label{de_m}
\delta\varepsilon_m=m(t)\delta(\mathbf{r}),
\end{equation}
where $\delta(\mathbf{r})$ is the Dirac invariant $\delta$-function defined with respect to the unperturbed metric and understood in the sense of the integral relation
\begin{eqnarray}\label{d(r)}
\int d^3V \delta(\mathbf{r})=4\pi a^3\int\delta(r)r^2dr=1,
\end{eqnarray}
so:
\begin{equation}\label{int_de}
\int d^3V \delta\varepsilon_m=m(t).
\end{equation}

Consider the singular equation with respect to $\nu(r,t)$ corresponding to the operator $\Delta_r$
\begin{equation}\label{sing_eq}
\frac{1}{r^2}\frac{\partial}{\partial r}\biggl(r^2\frac{\partial \nu}{\partial r}\biggr)=8\pi a^2 m(t)\delta(r).
\end{equation}
We select the \emph{singular part of the solution} in \eqref{sing_eq}, eq., assuming that the corresponding singular part is also contained in the particle energy density perturbation $\delta\varepsilon_p$ on the right side of \eqref{Einst_de} equation. Multiplying both parts of \eqref{sing_eq} by $r^2dr$ and integrating the resulting relation, we obtain by definition \eqref{d(r)}
\begin{equation}\label{dnu=}
r^2\frac{\partial \nu}{\partial r}=\frac{2m(t)}{a(t)}.
\end{equation}
Integrating \eqref{dnu=}, we finally find \emph{ for the singular part of the solution}
\begin{equation}\label{nu=}
\nu(r,t)=-\frac{2m(t)}{a(\eta)r}.
\end{equation}
Thus, we have a relation similar to the well-known one (see, for example, \cite{Land_Field}):
\begin{eqnarray}\label{Delta(m/r)=}
\frac{1}{r^2}\frac{\partial}{\partial r}r^2\frac{\partial}{\partial r}\biggl(-\frac{m(t)}{r}\biggr)=4\pi a^3m(t)\delta(r)
\quad (\equiv 4\pi a^3 \delta\varepsilon_m).
\end{eqnarray}

So, to isolate the particle-like singular part of the solution, we introduce a new scalar function $\rho(r,t)$ \cite{YuPhys1_08}
\begin{equation}\label{nu=1-2}
\nu(r,t)=-\lambda(r,t)=2\frac{\rho(r,t)-m(t)}{a(t)r}\equiv -\frac{2\mu(r,t)}{a(t)r},
\end{equation}
which is not singular at the origin:
\begin{equation}\label{rho/r}
\rho(0,t)=0;\quad  \lim_{r\to 0}r\frac{\partial\rho(r,t)}{\partial r}=0; \quad  \biggl|\lim_{r\to 0}\frac{\rho}{r}\biggr|< \infty.
\end{equation}

We select out the singular part in the equation for the perturbation of the scalar field \eqref{Eq_varphi}, setting similarly \eqref{de_m} and \eqref{nu=1-2}
\begin{equation}\label{s_q}
\delta\sigma_q=q(t)\delta(r)
\end{equation}
and
\begin{eqnarray}\label{phi=1-2}
\varphi(r,t)=2\frac{\chi(r,t)-q(t)}{a(t)r}\equiv \frac{2\phi(r,t)}{a(t)r},\qquad\\
\label{phi/r}
\chi(0,t)=0;\; \lim_{r\to 0}r\frac{\partial \chi(r,t)}{\partial r}=0;\biggl|\lim_{r\to 0}\frac{\chi}{r}\biggr|< \infty.
\end{eqnarray}
where $q(t)$ is a point scalar charge.
Thus,
\begin{equation}\label{nu(0)}
\left.\nu(r,t)\right|_{r\to0}\approx -\frac{2m(t)}{a(t)r},\; \left.\varphi(r,t)\right|_{r\to0}\approx -\frac{2q(t)}{a(t)r}.
\end{equation}

After substitution \eqref{nu=1-2}, the linearized Einstein equations \eqref{Einst_dp} -- \eqref{Einst_de} will take the following form:
\begin{eqnarray}\label{Einst_dpa}
\ddot{\mu}+2H\dot{\mu}-4\pi(\varepsilon+ p)_p\mu+Z\dot{\phi}-HZ\phi\nonumber\\ 
-\Phi(m_0^2-\alpha\Phi^2)\phi=-4\pi ar\delta p_p;\\
\label{Einst_dea}
-\frac{1}{a^2}\mu''+3H\dot{\mu}-\frac{1}{2}Z^2\mu-Z\dot{\phi}+HZ\phi-\Phi(m_0^2-\alpha\Phi^2)\phi =4\pi ar \delta\varepsilon_p;\\ 
\label{va=}
\upsilon=\frac{\partial}{\partial r}\biggl(\frac{1}{r}\frac{\dot{\mu}+H\phi}{4\pi a^3(\varepsilon+p)_p}\biggr).
\end{eqnarray}

Carrying out similar calculations for perturbations of the scalar field equation, taking into account \eqref{dZ/dt}, we obtain:
\begin{eqnarray}\label{EQ_phia}
\ddot{\phi}+H\dot{\phi}-\frac{1}{a^2}\phi''-\frac{2}{3}[2\pi(\varepsilon-3p)+\Lambda]\phi+2Z\dot{\mu}\nonumber\\
-[2HZ+\Phi(m_0^2-\alpha\Phi^2)+8\pi\sigma]\mu=-4\pi ar\delta\sigma.
\end{eqnarray}

When calculating the coefficients in the equations \eqref{Einst_dpa}, \eqref{va=} and \eqref{EQ_phia}, -- $(\varepsilon+ p)_p$ and $(\varepsilon-3p)$, defined with respect to background solutions , you must keep in mind the relations \eqref{2_3c}, \eqref{2_3a_2}, \eqref{{e+p}_p}, \eqref{Es} and \eqref{Ps}. Thus, we find, for example, for the coefficient in the equation \eqref{EQ_phi}:
\[2\pi(\varepsilon-3p)=-\frac{Z^2}{4}-\frac{1}{4\alpha}(m_0^2-\alpha\Phi^2)^2+\frac{e^4\Phi^4}{\pi}F_1(\psi).\]

In the right parts of the equations \eqref{Einst_dpa}, \eqref{Einst_dea} and \eqref{EQ_phia} it is necessary to substitute the expressions for perturbations of scalar densities $\delta p_p$ \eqref{dp}, $\delta\varepsilon_p$ \eqref{de} and $\delta\sigma$ \eqref{dsigma}, respectively. These perturbations of scalar densities, in turn, are completely determined by the linear functions of perturbations $\delta_z$ and $\varphi$. Substituting \eqref{phi=1-2} into these values and
\begin{equation}\label{delta_z->}
\delta_z(r,t)=2\frac{d_z(r,t)}{a(t)r},
\end{equation}
we obtain a closed system of three corresponding linear differential equations \eqref{Einst_dpa}, \eqref{Einst_dea} and \eqref{EQ_phia} with respect to three independent non-singular functions $\mu(r,t)$, $\phi(r,t)$ and $\tilde{\delta}_z(r,t)$. Note that in this case the function $\tilde{\delta}_z(r,t)$ enters these equations algebraically. Therefore, this function can be expressed in terms of the functions $\mu(r,t)$ and $\phi(r,t)$ and their derivatives. The easiest way to do this is with the \eqref{Einst_dpa} equation, because it does not contain derivatives with respect to the radial variable.

According to \eqref{nu=1-2} and \eqref{phi=1-2} near the singularity $r=0$, the perturbed metric \eqref{metric_pert} in the approximation linear in smallness of perturbations can be written as a metric conformal to the Schwarzschild metric with physical radius $R=ar$ \eqref{R=}
%
\begin{equation}\label{shvarc}
ds^2\backsimeq  \biggl(1-\frac{2m(t)}{a(t)r}\biggr)dt^2 -\frac{dr^2+r^2(d\Omega^2+\sin^2\Omega d\varphi^2)}{\displaystyle 1-\frac{2m(t)}{a(t)r}}.
\end{equation}
%
In this regard, we will henceforth call solutions with $\mu(t)\not=0,\ \phi(t)\not=0$ \emph{scalar black holes (SBH)}\label{_SBH}, keeping in mind that the solution of type \eqref{shvarc} is a \emph{linear model} of a black hole (see \cite{Yu_BH_19}). In this case, the evolution of the mass and charge of SBH reflects the ongoing processes of accretion and scattering of the fermionic and scalar components of matter.

Note that according to \eqref{dF-drho-du}, \eqref{nu=1-2} and \eqref{phi=1-2}, for the validity of the linear approximation in the field equations, the following conditions must be satisfied
\begin{equation}\label{nu,phi<<1}
\nu(r,t)\ll \frac{1}{2}a(t)r,\qquad \phi(r,t)\ll \frac{1}{2}a(t)r.
\end{equation}
This implies that for $r\to0$ the perturbations tend to infinity, and the condition of small perturbations is violated. However, this is a purely formal conclusion. Indeed, as it follows from \eqref{shvarc}, on the contrary, near the singularity, the linear solution with $m(t)\not=0$ exactly coincides locally with the exact spherically symmetric solution of the Einstein equations in the medium (see, for example, \ cite{Sing}, \cite{Land_Field}) under substitutions $ar\to R,\ m(t)\to \mu(r,t)$.
On the one hand, this is a consequence of the law of conservation of total energy, and on the other hand, it is a well-known fact that the consistent theory of approximations, taking into account approximations of the energy of the gravitational field in higher orders, gives the exact Schwarzschild solution from the classical Poisson equation with a singular source. Therefore, we should not be confused by the appearance of terms of the $1/r$ type in the expressions for the perturbation modes. Note that from a formal point of view, we would not have the right to consider in the theory of gravity any point sources, both of the gravitational field and of other fields, including even elementary particles (see \cite{TMF_21} in this connection).


%
\subsection{System of equations for nonsingular functions\label{subs_eqs}}
So, using the equations \eqref{Einst_dpa} and \eqref{dp} we find the function $d_z(r,t)$:
\begin{eqnarray}\label{tile_delta_z}
d_z(r,t)=-\frac{3\pi}{8e^4\Phi^4}\frac{\sqrt{1+\psi^2}}{\psi^5}\ddot{\mu}+2H\dot{\mu}+Z\dot{\phi}+\frac{1+\psi^2}{2\psi^4}\mu +\biggl[\frac{\sqrt{1+\psi^2}}{2\Phi\psi^4}F_1(\psi)\\
+\frac{3\pi}{8e^4\Phi^4}\frac{\sqrt{1+\psi^2}}{\psi^5}\bigl(\Phi(m_0^2-\alpha\Phi^2)+ HZ\bigr)\biggr]\phi.\nonumber
\end{eqnarray}
Using this relation in the expressions for perturbations of macroscopic scalars $\delta\varepsilon_p$ \eqref{de} and $\delta\sigma$ \eqref{dsigma} and then substituting the resulting expressions into the equations \eqref{Einst_dea} and \eqref{EQ_phia}, we obtain a closed system of homogeneous second-order linear differential equations with respect to the functions $\mu(r,t)$ and $\phi(r,t)$:
\begin{eqnarray}\label{EQ_mu}
\frac{1}{a^2}\mu''-3\frac{(1+\psi^2)}{\psi^2}\ddot{\mu}-3H\frac{2+3\psi^2}{\psi^2}\dot{\mu}-Z\frac{3+2\psi^2}{\psi^2}\dot{\phi}+\mathrm{L}_{\varepsilon m}(t)\mu+\mathrm{L}_{\varepsilon q}(t)\phi=0;\\
\label{EQ_phi}
\ddot{\phi}+\biggl(H-\frac{3}{2\Phi\psi^2}\biggr)\dot{\phi}-\frac{1}{a^2}\phi'' -\frac{3}{2\Phi\psi^2}\ddot{\mu}
+\biggl(\frac{3H}{\Phi\psi^2}-2Z\biggr)\dot{\mu}+\mathrm{L}_{\sigma m}(t)\mu+\mathrm{L}_{\sigma q}(t)\phi=0.
\end{eqnarray}

Functions $\mathrm{L}_{\alpha\beta}(t)$ are introduced into \eqref{EQ_mu} -- \eqref{EQ_phi}, which are determined by background solutions $\{\xi(t)$, $H(t )$, $\Phi(t)$, $Z(t)\}$:
\begin{eqnarray}\label{L_em}
\mathrm{L}_{\varepsilon m}(t)=\frac{Z^2}{2}+\frac{4}{\pi}e^4\Phi^4\frac{(1+\psi^2)^{3/2}}{\psi};\nonumber\\
\label{L_eq}
\mathrm{L}_{\varepsilon q}(t)=HZ\frac{3+2\psi^2}{\psi^2}+\frac{3+4\psi^2}{\psi^2}\Phi(m_0^2-\alpha\Phi^2)+4e^4\Phi^3\frac{1+2\psi^2}{\pi\psi^2}F_1(\psi);\nonumber
\end{eqnarray}
%
\begin{eqnarray}\label{L_sm}
\mathrm{L}_{\sigma m}(t)=-2HZ-\Phi(m_0^2-\alpha\Phi^2)+\frac{2e^4\Phi^3}{\pi\psi}\bigl(\sqrt{1+\psi^2}-2\psi F_1(\psi)\bigr);\nonumber
\end{eqnarray}
\begin{eqnarray}
\label{L_sq}
\mathrm{L}_{\sigma q}(t)=\frac{3HZ}{2\Phi\psi^2}-\frac{2}{3}\Lambda+\frac{3}{2\psi^2}(m_0^2-\alpha\Phi^2)+\frac{1}{6\alpha}(m_0^2-\Phi^2)^2)+\frac{Z^2}{6}+\frac{8e^4\Phi^2\psi^3}{\pi\sqrt{1+\psi^2}}\nonumber\\
+\frac{2e^4\Phi^2}{\pi\psi^2}F_1(\psi)\biggl(1-6\psi^2-\frac{\Phi^2\psi^2}{3}\biggr).
\nonumber
\end{eqnarray}
It is also necessary to substitute the value of the function $\psi$ from \eqref{psi(eta)} into these relations.

Thus, the problem has been reduced to a system of two linear homogeneous second-order partial differential equations with respect to two functions $\mu(r,t)$ and $\phi(r,t)$: \eqref{EQ_mu} -- \eqref{EQ_phi }.
As can be seen, the coefficients of these equations are described by rather cumbersome expressions for the basic background functions of the $\{\xi(t),H(t),\Phi(t),Z(t)\}$ model. However, an important feature of these equations is the dependence of their coefficients only on time.

Before studying these equations, we note their important property, which allows us to significantly simplify the problem. Consider a system of homogeneous linear differential equations with respect to a system of vector functions of two variables $\mathbf{\mathbf{U}}(x,t)$\footnote{The order of the equations and the number of functions are not essential.}:
\begin{equation}\label{Sys0}
[\mathbf{X}(x)+\mathbf{T}(t)]\mathbf{W}(x,t)=0,
\end{equation}
where $\mathbf{X}(x)$ and $\mathbf{T}(t)$ are linear matrix operators, and
\begin{equation}\label{XU=0}
\mathbf{X}(x)\mathbf{W}(t)=0.
\end{equation}
It is easy to see that the \eqref{EQ_mu} -- \eqref{EQ_phi} system of equations refers to systems of the \eqref{Sys0} type with the condition \eqref{XU=0}.
Let's consider solutions of equations \eqref{Sys0} of the form:
\begin{equation}\label{W=a}
\mathbf{W}=\mathbf{U(}t)+\mathbf{V}(x,t),\quad \frac{\partial}{\partial x} \mathbf{V}\not=\mathbf{0}.
\end{equation}
Substituting \eqref{W=a} into \eqref{Sys0}, taking into account the condition \eqref{XU=0} we get:
\begin{equation}\label{U+V}
[\mathbf{X}(x)+\mathbf{T}(t)]\mathbf{V}(x,t)+\mathbf{T}(t)\mathbf{U}(t)=0.
\end{equation}
Let us write the condition that the second term of this equation depends only on the variable $t$:
\begin{equation}\label{Eq0_U(t)}
\mathbf{T}(t)\mathbf{U}(t)=\mathbf{f}(t),
\end{equation}
where $\mathbf{f}(t)$ is an arbitrary vector function of the argument $t$. But then \eqref{U+V} implies
\begin{equation}\label{Eq0_V(r,t)}
[\mathbf{X}(x)+\mathbf{T}(t)]\mathbf{V}(x,t)=-\mathbf{f}(t).
\end{equation}
Since the equation \eqref{Eq0_U(t)} is linear, its general solution is the sum of the general solution of the corresponding homogeneous equation, $U_0(t)$
\begin{equation}\label{U0}
\mathbf{T}(t)\mathbf{U}_0(t)=0
\end{equation}
and any particular solution of the inhomogeneous equation, $\mathbf{U}_1(t)$:
\begin{equation}\label{Eq1_U(t)}
\mathbf{T}(t)\mathbf{U}_1(t)=\mathbf{f}(t),
\end{equation}
so that:
\begin{equation}\label{U=}
\mathbf{U}(t)=\mathbf{U}_0(t)+\mathbf{U}_1(t).
\end{equation}

Further, the equation \eqref{Eq0_V(r,t)} is also linear and homogeneous. Let $\mathbf{V}_1(x,t)$ be the general solution of the corresponding homogeneous equation
\begin{equation}\label{V0}
[\mathbf{X}(x)+\mathbf{T}(t)]\mathbf{V}_0(x,t)=0.
\end{equation}
Due to the linearity of the inhomogeneous equation \eqref{Eq0_V(r,t)}, its particular solution is $\mathbf{V}_1(x,t)$ $=-\mathbf{U}_0(t)$. But then
\begin{equation}\label{V=}
\mathbf{V}(t)=\mathbf{V}_0(x,t)-\mathbf{U}_1(t).
\end{equation}
Thus, the following statement is true:
\begin{stat}
\emph{The general solution of the \eqref{U+V} equation is the sum of the solutions of the homogeneous equations \eqref{U0} and \eqref{V0}:}
\begin{equation}\label{W=}
\mathbf{W}(x,t)=\mathbf{V}_0(x,t)+\mathbf{U}_0(t).
\end{equation}
\end{stat}

Proved property, taking into account the relations \eqref{nu=1-2} and \eqref{phi=1-2}
\[\mu(r,t)=\rho(r,t)-m(t),\; \phi(r,t)=\chi(r,t)-q(t)\]
allows writing the system \eqref{EQ_mu} -- \eqref{EQ_phi} as a system of four linear homogeneous equations, of which the first two equations form an independent subsystem of ordinary homogeneous differential equations with respect to the source functions $m(t)$ and $q(t) $ (\eqref{EQ_m} -- \eqref{EQ_q})
\begin{eqnarray}\label{EQ_m}
-3\frac{(1+\psi^2)}{\psi^2}\ddot{m}-3H\frac{2+3\psi^2}{\psi^2}\dot{m}-Z\frac{3+2\psi^2}{\psi^2}\dot{q}+\mathrm{L}_{\varepsilon m}(t)m+\mathrm{L}_{\varepsilon q}(t)qi=0;\\
\label{EQ_q}
\ddot{q}+\biggl(H-\frac{3}{2\Phi \psi^2}\biggr)\dot{q} -\frac{3}{2\Phi\psi^2}\ddot{m}+\biggl(\frac{3H}{\Phi\psi^2}-2Z\biggr)\dot{m}+\mathrm{L}_{\sigma m}(t)m+\mathrm{L}_{\sigma q}(t)q=0.
\end{eqnarray}
and the remaining two equations constitute an independent subsystem of homogeneous partial differential equations with respect to the functions $\rho(r,t)$ and $\chi(r,t)$ (\eqref{EQ_rho} -- \eqref{EQ_chi}) satisfying the conditions \eqref{rho/r} and \eqref{phi/r} at origin
\begin{eqnarray}\label{EQ_rho}
\frac{1}{a^2}\rho''-3\frac{(1+\psi^2)}{\psi^2}\ddot{\rho}-3H\frac{2+3\psi^2}{\psi^2}\dot{\rho}-Z\frac{3+2\psi^2}{\psi^2}\dot{\chi}+\mathrm{L}_{\varepsilon m}(t)\rho+\mathrm{L}_{\varepsilon q}(t)\chi=0;\\
\label{EQ_chi}
\ddot{\chi}+\biggl(H-\frac{3}{2\Phi\psi^2}\biggr)\dot{\chi}-\frac{1}{a^2}\chi'' -\frac{3}{2\Phi\psi^2}\ddot{\rho}
+\biggl(\frac{3H}{\Phi\psi^2}-2Z\biggr)\dot{\rho}+\mathrm{L}_{\sigma m}(t)\rho+\mathrm{L}_{\sigma q}(t)\chi=0.
\end{eqnarray}

The equations \eqref{EQ_m} and \eqref{EQ_q} will be called evolution equations for the mass and scalar charge of a singular source. Since these equations are linear and homogeneous, then under zero initial conditions
$m(t_0)=0$; $q(t_0)=0$ they have only the trivial solution $m(t)=0,\ q(t)=0$. If at least one of the initial conditions is not zero, then $m(t)\not=0,\ q(t)\not=0$, which is necessary for the formation of a scalar black hole.
Two subsystems of equations \eqref{EQ_m} -- \eqref{EQ_q} and \eqref{EQ_rho} and \eqref{EQ_chi}, as we noted, are independent, but their solutions can be related by the boundary conditions of the Cauchy problem.
\subsection{Shortwave limit}
We study disturbances in the shortwave sector based on the equations \eqref{EQ_rho} -- \eqref{EQ_chi}, assuming
\begin{eqnarray}\label{eikonal}
\displaystyle \rho(r,t)=\tilde{\rho}(t)\mathrm{e}^{inar+i\int u(t)dt};&
 \chi(r,t)=\tilde{\chi}(t)\mathrm{e}^{ianr+i\int u(t)dt}; & (na\gg1,\dot{a}r\ll 1),
\end{eqnarray}
where $\tilde{\rho}(t)$, $\tilde{\chi}(t)$ and $u(t)$ are slowly varying functions of the time variable $t$, and $u(t)\sim n$ is the eikonal function. Substituting \eqref{eikonal} into \eqref{EQ_rho} -- \eqref{EQ_chi},
in the zero WKB approximation, we obtain two oscillation modes:
\begin{eqnarray}\label{WKB}
\rho:& u=\pm n\frac{\psi}{\sqrt{3(1+\psi^2)}},& \Rightarrow v_f =\frac{|u|}{n}\leqslant\frac{1}{\sqrt{3}};\\
\chi:& u=\pm n,& \Rightarrow v_f =\frac{|u|}{n}=1.
\end{eqnarray}
These independent oscillation modes correspond to two pairs of diverging and converging waves propagating with the phase velocity $v_f=u/n$, and the waves corresponding to perturbations of the gravitational potential propagate at the speed of sound in the medium of fermions, and the waves corresponding to perturbations of the scalar field potential propagate at the speed of light. In accordance with \eqref{eikonal}, \eqref{nu=1-2} and \eqref{phi=1-2}, the amplitudes of these waves change according to the law:
\[\tilde{\nu}\sim \frac{1}{a(t)r},\quad \tilde{\varphi}\sim \frac{1}{a(t)r}.\]
These high-frequency modes are undamped, the imaginary parts of the eikonal function $u(r,t)$, corresponding to both damping and growth of oscillations, appear when the higher-order terms of the WKB approximation are taken into account (see \cite{Yu_GC_3_22}).

\section{Solution for localized perturbations}

\subsection{Localized disturbances}
Let us consider the case of spherical perturbations localized in a sphere of radius $r_0$. The localization of disturbances inside the sphere means that at the boundary of the sphere, the disturbances and their derivatives with respect to the radial variable vanish, i.e.,
boundary conditions are satisfied
\begin{eqnarray}\label{bound_0}
\nu(r_0,0)=0; \quad {\displaystyle \left.\frac{d\nu(r,0)}{dr}\right|_{r=r_0}=0};\quad
\phi(r_0,0)=0; \quad {\displaystyle  \left.\frac{d\phi(r,0)}{dr}\right|_{r=r_0}=0}.
\end{eqnarray}
It is easy to see that, due to the definitions \eqref{nu=1-2} -- \eqref{phi=1-2}, the boundary conditions \eqref{bound_0} can be rewritten as:
\begin{eqnarray}\label{bound_1}
\rho(r_0,0)=\mu(0); \quad {\displaystyle \left.\frac{d\rho(r,0)}{dr}\right|_{r=r_0}=0}; &
\chi(r_0,0)=q(0); \quad {\displaystyle  \left.\frac{d\chi(r,0)}{dr}\right|_{r=r_0}=0}.
\end{eqnarray}

\TextFigReg{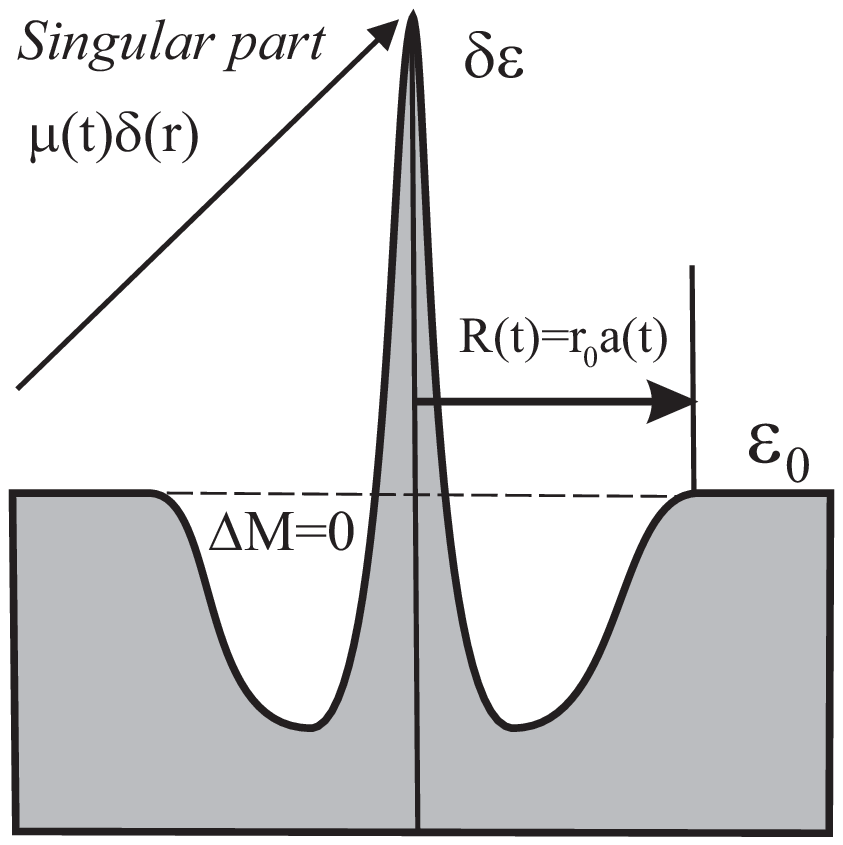}{7}{\label{ris1}Schematic representation of the structure of a localized perturbation}{7}{\quad Note that the localization of perturbations means that the change in the total mass $M(V)$ and the charge $Q(V)$ of the localization region $V$ are equal to zero. Otherwise, due to the conservation, for example, of mass, the space outside the perturbation region would not be homogeneous, i.e., the solution of the Einstein equations would not give the Friedmann Universe.

\quad Thus, in localized perturbations the matter density is only redistributed due to the Friedmann matter \emph{inside the localization sphere}. This means that if a singular mass occurs somewhere, then in the surrounding sphere of perturbation localization, the perturbation of the energy density must be negative (see Fig. \ref{ris1}). The foregoing applies entirely to the localization of the scalar charge density.}


Assuming that in a finite neighborhood $r\in[0,r_0)$, in which the perturbation of the metric is localized, the potential functions $\rho(r,\eta)$ \eqref{nu=1-2} and $\chi(r ,\eta)$ \eqref{phi=1-2} satisfying the conditions \eqref{rho/r} and \eqref{phi/r} belong to the class $\mathrm{C}^\infty$, we represent the solutions of the equations \eqref{EQ_rho} -- \eqref{EQ_chi} in the form of power series\footnote{It is the localization of perturbations that makes it possible to expand perturbations into power series.} that do not contain terms with zero degree due to the conditions \eqref{rho/r} $\rho (0,t)=0$ and \eqref{phi/r} $\chi(0,t)=0$;
\begin{equation}\label{rho_chi_sol_series}
\rho(r,t)=\sum\limits_{n=1}^\infty \rho_n(t)r^n,\; \chi(r,t)=\sum\limits_{n=1}^\infty \chi_n(t)r^n.
\end{equation}
In this case, obviously, the initial values of the potential functions $\rho(r,0)$ and $\chi(r,0)$ should look like:
\begin{equation}\label{rho_chi_IC_series0}
\rho(r,0)=\sum\limits_{n=1}^\infty c_n r^n, \qquad \chi(r,0)=\sum\limits_{n=1}^\infty d_n r^n,
\end{equation}
so that the initial conditions on the functions $\rho_n(t)$ and $\chi_n(t)$
\begin{equation}\label{rho_chi_IC}
\rho_n(0)=c_n, \qquad \chi_n(0)=d_n.
\end{equation}
\subsection{Equations for localized disturbances}
Substituting the expansions \eqref{rho_chi_sol_series} into the system of equations \eqref{EQ_rho} -- \eqref{EQ_chi} and equating the terms with the same powers of $r$, we immediately obtain
\begin{equation}\label{x_2m=0}
\rho_{2m}=0;\qquad \chi_{2m}=0, \qquad (m=\overline{0,\infty}).
\end{equation}
Thus,
\begin{eqnarray}\label{rho_chi_sol_series1}
\rho(r,t)=\sum\limits_{k=0}^\infty \rho_{2k+1}(t)r^{2k+1}, &
\chi(r,t)=\sum\limits_{k=0}^\infty \chi_{2k+1}(t)r^{2k+1},
\end{eqnarray}
-- as we noted above, the potential functions $\rho(r,t)$ and $\chi(r,t)$ are odd functions of $r$.

Due to the boundary conditions at the zero time \eqref{bound_0} and \eqref{rho_chi_IC} the following conditions must be satisfied for the coefficients $c_n,d_n$:
\begin{eqnarray}\label{Inits_cd}
\sum\limits_{k=0}^\infty c_{2k+1}(0)r_0^{2k+1}=m(0); & \sum\limits_{k=0}^\infty d_{2k+1}(0)r_0^{2k+1}=q(0);\nonumber\\
\sum\limits_{k=0}^\infty kc_{2k+1}(0)r_0^{2k+1}=-\frac{m(0)}{2}; & \sum\limits_{k=0}^\infty kd_{2k+1}(0)r_0^{2k+1}=-\frac{q(0)}{2}.
\end{eqnarray}

For the expansion coefficients \eqref{rho_chi_sol_series1}, which are functions of time, we obtain recurrent equations:

\begin{eqnarray}
\label{rho_2m+3}
(2k+3)(2k+2)\frac{\rho_{2k+3}}{a^2(t)}= 3\frac{1+\psi^2}{\psi^2}\ddot{\rho}_{2k+1}+3H\frac{2+3\psi^2}{\psi^2}\dot{\rho}_{2k+1}-\mathrm{L}_{\varepsilon m}(t)\rho_{2k+1}\nonumber\\
+Z\frac{3+2\psi^2}{\psi^2}\dot{\chi}_{2k+1}-\mathrm{L}_{\varepsilon q}(t)\chi_{2k+1};\\
\label{chi_2m+3}
(2k+3)(2k+2)\frac{\chi_{2k+3}}{a^2(t)}= \ddot{\chi}_{2k+1} +\biggl(H-\frac{3}{2\Phi\psi^2}\biggr)\dot{\chi}_{2k+1}- \frac{3}{2\Phi\psi^2}\ddot{\rho}_{2k+1}\nonumber\\
+\biggl(\frac{3H}{\Phi\psi^2}-2Z\biggr)\dot{\rho}_{2k+1}+\mathrm{L}_{\sigma q}(t)\chi_{2k+1}+\mathrm{L}_{\sigma m}(t)\rho_{2m+1}.
\end{eqnarray}

Depending on the initial conditions, the expansion \eqref{rho_chi_sol_series1} can be terminated at any $n=n_0\geqslant 3$. Let it be $n_0=2k_0+1$. Then, since $\rho_{2k_0+3}=0$, $\chi_{2k_0+3}=0$, for the coefficients of the last terms of expansions \eqref{rho_chi_sol_series1} we obtain from \eqref{rho_2m+3} -- \eqref{chi_2m+3} equations:
\begin{eqnarray}
\label{EQrho_2m+1}
\!\!\! 3\frac{1+\psi^2}{\psi^2}\ddot{\rho}_{2k_0+1}+3H\frac{2+3\psi^2}{\psi^2}\dot{\rho}_{2k_0+1}
+Z\frac{3+2\psi^2}{\psi^2}\dot{\chi}_{2k_0+1}-\mathrm{L}_{\varepsilon m}(t)\rho_{2k_0+1}
-\mathrm{L}_{\varepsilon q}(t)\chi_{2k_0+1}=0;\\
\label{EQchi_2m+1}
\ddot{\chi}_{2k_0+1}+\biggl(H-\frac{3}{2\Phi\psi^2}\biggr)\dot{\chi}_{2k_0+1}-\frac{3}{2\Phi\psi^2}\ddot{\rho}_{2k_0+1}+\biggl(\frac{3H}{\Phi\psi^2}-2Z\biggr)\dot{\rho}+\mathrm{L}_{\sigma q}(t)\chi_{2k_0+1}+\mathrm{L}_{\sigma m}(t)\rho_{2k_0+1}=0.
\end{eqnarray}
Since, firstly, the system of equations \eqref{EQrho_2m+1} -- \eqref{EQchi_2m+1} is a homogeneous system of ordinary linear differential equations, and secondly, it coincides with the system
\eqref{EQ_m} -- \eqref{EQ_q} equations, then its solutions that satisfy the initial conditions \eqref{rho_chi_IC} can only be the following:
\begin{eqnarray}\label{rho,chi_2m+1=}
\rho_{2k_0+1}(t)=c_{2k_0+1}\frac{m(t)}{m(0)}; & \chi_{2k_0+1}(t)=d_{2k_0+1}\frac{q(t)}{q(0)}.
\end{eqnarray}

Substituting then the solutions \eqref{rho,chi_2m+1=} into the left-hand side of the equations \eqref{rho_2m+3} -- \eqref{chi_2m+3} (assuming $k_0\to k_0-1$), we obtain a system of differential equations for the functions $\rho_{2k_0-1}$ and $\chi_{2k_0-1}$:

\begin{eqnarray}
\label{rho_2m+1}
3\frac{1+\psi^2}{\psi^2}\ddot{\rho}_{2k_0-1}+3H\frac{2+3\psi^2}{\psi^2}\dot{\rho}_{2k_0-1}+Z\frac{3+2\psi^2}{\psi^2}\dot{\chi}_{2k_0-1}-\mathrm{L}_{\varepsilon m}(t)\rho_{2k_0-1}-\mathrm{L}_{\varepsilon q}(t)\chi_{2k_0-1}\nonumber\\
=\frac{2k_0(2k_0+1)c_{2k_0+1}}{a^2(t)}\frac{m(t)}{m(0)};
\end{eqnarray}
\begin{eqnarray}
\label{chi_2m+1}
\ddot{\chi}_{2k_0-1}+\biggl(H-\frac{3}{2\Phi\psi^2}\biggr)\dot{\chi}_{2k_0-1}-\frac{3}{2\Phi\psi^2}\ddot{\rho}_{2k_0-1}
+\biggl(\frac{3H}{\Phi\psi^2}-2Z\biggr)\dot{\rho}_{2k_0-1}+\mathrm{L}_{\sigma q}(t)\chi_{2k_0-1}+\mathrm{L}_{\sigma m}(t)\rho_{2k_0-1}\nonumber\\
=\frac{2k_0(2k_0+1)d_{2k_0+1}}{a^2(t)}\frac{q(t)}{q(0)}.\hskip 12pt
\end{eqnarray}
These equations must be solved with initial conditions \eqref{rho_chi_IC}. Note that the left parts of the equations \eqref{rho_2m+1} -- \eqref{chi_2m+1} coincide with the left parts of the homogeneous system of equations \eqref{EQ_m} -- \eqref{EQ_q} that determine the evolution of the mass $m(t) $ and charge $q(t)$. The solutions of the equations \eqref{rho_2m+1} -- \eqref{chi_2m+1}, in turn, determine the equations for the functions $\rho_{2m_0-3}$ and $\chi_{2m_0-3}$, etc.. Thus, we have obtained a recursive procedure for finding all coefficients of the series \eqref{rho_chi_sol_series1}, i.e., we have obtained \emph{exact solution of equations for localized perturbations} up to integration of a system of ordinary linear inhomogeneous differential equations of the second order \eqref{rho_2m+1} -- \eqref{chi_2m+1}.

Thus, as we noted at the end of the \ref{subs_eqs} section, the perturbation pairs $\{\rho(r,t), \chi(r,t)\}$ and $\{m(t), q(t )\}$, indeed, turn out to be related through the initial-boundary conditions. However, the perturbations
$\{m(t), q(t)\}$, which are responsible for the evolution of the singular mass and charge, remain independent of the other pair of perturbations, and depend only on the initial conditions. This property of the evolution of spherical perturbations is very important.

\subsection{Perturbations in the form of a polynomial of the third degree in a radial variable}
Let us consider the simplest form of a localized perturbation corresponding to a polynomial of the third degree. In accordance with the formulas \eqref{rho_chi_sol_series1} in the case of $k_0=3$, the corresponding expansions have the form
\begin{eqnarray}\label{cub_0}
\rho(r,t)=\rho_1(t) r+\rho_3(t)r^3; \quad 
\chi(r,t)=\chi_1(t) r+\chi_3(t)r^3.
\end{eqnarray}
Substituting \eqref{cub_0} into the boundary conditions \eqref{bound_1}, we find the initial values of potential functions (or directly from \eqref{Inits_cd})
\begin{eqnarray}
\rho(r,0)=\frac{m(0)}{2}\biggl(3\frac{r}{r_0}-\frac{r^3}{r^3_0}\biggr)\Rightarrow
\nu(r,0)=m(0)\biggl(\frac{3}{r_0}-\frac{r^2}{r^3_0}-\frac{2}{r}\biggr);\nonumber
\end{eqnarray}
\begin{eqnarray}\label{IC_rho-chi}
\chi(r,0)=\frac{q(0)}{2}\biggl(3\frac{r}{r_0}-\frac{r^3}{r^3_0}\biggr)\Rightarrow
\varphi(r,0)=q(0)\biggl(\frac{3}{r_0}-\frac{r^2}{r^3_0}-\frac{2}{r}\biggr).
\end{eqnarray}
Here we have taken into account that according to \eqref{xi(0)} $a(0)=1$. In accordance with the decomposition of \eqref{rho_chi_sol_series1} and the initial conditions \eqref{IC_rho-chi}, we obtain the initial conditions for the functions $\rho_1(\eta)$, $\chi_1(\eta)$:
\begin{equation}\label{IC_rho_1-chi_1}
\rho_1(0)=\frac{3m(0)}{2r_0}\equiv \rho_1^0;\; \chi_1(0)=\frac{3q(0)}{2r_0}\equiv \chi_1^0.
\end{equation}

Thus, according to \eqref{rho,chi_2m+1=}, \eqref{rho_2m+1} and \eqref{chi_2m+1}, we find, firstly, --
\begin{equation}\label{rho,chi_3=}
\rho_{3}(t)=-\frac{m(t)}{2r^3_0};\;\chi_{3}(t)=-\frac{q(t)}{2r^3_0}.
\end{equation}
Second, substituting \eqref{rho,chi_3=} into \eqref{cub_0}, we find expressions for the functions $\rho(r,t)$ and $\chi(r,t)$ in terms of the evolution functions $m(t )$, $q(t)$, $\rho_1(t)$ and $\chi_1(t)$
\begin{eqnarray}\label{cub_1}
\rho(r,t)=\rho_1(t) r-m(t)\frac{r^3}{2r^3_0}; 
\quad \chi(r,t)=\chi_1(t) r-q(t)\frac{r^3}{2r^3_0}.
\end{eqnarray}

Substituting the solutions \eqref{rho,chi_3=} into the right-hand side of the equations \eqref{rho_2m+1} -- \eqref{chi_2m+1} (assuming $k_0=1$), we obtain a system of differential equations with respect to the functions
$\rho_1 (t)$ and $\chi_1(t)$:
\begin{eqnarray}
\label{rho_1}
3\frac{1+\psi^2}{\psi^2}\ddot{\rho}_{1}+3H\frac{2+3\psi^2}{\psi^2}\dot{\rho}_1+Z\frac{3+2\psi^2}{\psi^2}\dot{\chi}_1
-\mathrm{L}_{\varepsilon m}(t)\rho_{1}-\mathrm{L}_{\varepsilon q}(t)\chi_{1}=-\frac{3m(t)\mathrm{e}^{-2\xi(t)}}{2r^3_0};\\
\label{chi_1}
\ddot{\chi}_1+\biggl(H-\frac{3}{2\Phi\psi^2}\biggr)\dot{\chi}_1-\frac{3}{2\Phi\psi^2}\ddot{\rho}_1 
+ \biggl(\frac{3H}{\Phi\psi^2}-2Z\biggr)\dot{\rho}_1+\mathrm{L}_{\sigma q}(t)\chi_1+\mathrm{L}_{\sigma \mu}(t)\rho_1
=-\frac{3q(t)\mathrm{e}^{-2\xi(t)}}{2r^3_0}.
\end{eqnarray}
Thus, in the case of $n=3$, the problem is completely reduced to solving two systems of evolutionary equations: a system of two ordinary homogeneous linear equations of the second order \eqref{EQ_m} -- \eqref{EQ_q} and a system of two ordinary nonhomogeneous linear equations of the second order \eqref{rho_1} -- \eqref{chi_1}.

\subsection{Disturbance localization radius\label{r_section}}
The localization radius of a gravitational perturbation can be defined as the radius at which the radial derivative of the potential $\nu(r,\eta)$ vanishes.
Calculating this derivative, we find, taking into account the relation \eqref{rho,chi_3=}:
\begin{eqnarray}\label{r_m}
\nu'(r,t)=0\!\Rightarrow\!  r\rho'(r,t)-(\rho(r,t)-m(t))\! =0\Rightarrow -2\rho_3(t)r^3+m(t)=0\Rightarrow r_m(t)=r_0.
\end{eqnarray}
Thus, the localization radius of a gravitational perturbation, defined as the radius at which the gravitational attraction force of the perturbation disappears, does not change in isotropic coordinates $\{r,\theta,\varphi,\eta\}$.
Similarly, we can show that the localization radius of the perturbation of the scalar field coincides with $r_0$. Note that the physical radius is equal to $R=a(t)r$, so the physical radius of perturbation localization grows with the expansion of the Universe: $R_m=a(t)r_0$.

In this case, a direct calculation of the total gravitational mass of the perturbation $M(r,t)$ inside a sphere of radius $r_0$ in accordance with \eqref{shvarc} --
\begin{eqnarray}\label{M(r,t)}
\nu(r,t)=-\frac{2M(r,t)}{ar}\Rightarrow  M(r,t)=-\rho(r,t)+m(t)
\end{eqnarray}
does not result in a null value. Note that according to \eqref{cub_1} we get for \eqref{M(r,t)}
\begin{eqnarray}\label{nu_rt}
\nu(r,t)=\frac{2}{a(t)}\biggl(\rho_1(t)-\frac{m(t)r^2}{2r_0^3}-\frac{m(t)}{r}\biggr)
\end{eqnarray}
However, we have the right to add an arbitrary small gauge function of time $\delta\xi(t)$ to the right side \eqref{nu_rt} and transform the time variable \eqref{delta_t}. Let us use this invariant property of the metric and choose the function $\delta\xi(t)$ so that $M(r_0,t)=0$. This gives:
\begin{eqnarray}\label{dxi}
\delta\xi(t)=-\frac{2\rho_1(t)}{a(t)}+\frac{3m(t)}{r_0 a(t)}.
\end{eqnarray}

Now adding $\delta\xi(t)$ from \eqref{dxi} to the right hand side of \eqref{nu_rt}, we get $\nu_n$ for the renormalized potential of the gravitational perturbation
\begin{equation}\label{nu_cal}
\nu_n(r,t)=\frac{2m(t)}{a(t)}\biggl(\frac{3}{r_0}-\frac{r^2}{r_0}-\frac{2}{r}\biggl).
\end{equation}
It is obvious that this value of $\nu(r,t)$ automatically leads to the law of conservation of the total mass inside the disturbance localization region:
\begin{equation}\label{M=0}
M(r_0,t)=0.
\end{equation}
Note that we cannot demand similar conservation of the scalar charge, since the charge $q(t)$ is associated with the scalar charge density $\sigma$ \eqref{2}, which is not related to the conservation law
in contrast to the kinematic charge density \eqref{2c}. It suffices that at $r=r_0$ the gradient of the scalar potential vanishes, which leads to the force of scalar attraction equal to zero.

Note also that according to \eqref{shvarc}, the gravitational radius of the formed black hole can be estimated as\footnote{This is precisely the estimate, since the singular source is surrounded by matter.}
\begin{equation}\label{r_g}
r_g=2m(t).
\end{equation}
Since the mass of a black hole is formed entirely due to <<eating away>> matter from the localization region, its final mass cannot exceed the total mass of the localized perturbation. When this limit is reached, the growth of the mass of the Black Hole stops. Thus, the final mass of the Black Hole is determined by the initial parameters of the spherical perturbation at the time of its formation from unstable plane modes.

\subsection{Non-singular localized disturbances}
According to \eqref{nu=1-2} -- \eqref{phi=1-2}, non-singular perturbations correspond to zero values of the mass and charge functions:
\begin{equation}\label{nonsing}
m(t)\equiv0;\qquad q(t)\equiv 0.
\end{equation}
Then, firstly, from \eqref{rho,chi_2m+1=} it follows that the last members of the series \eqref{rho_chi_sol_series1} must look like:
\begin{eqnarray}\label{rho,chi_2m+1=new}
\rho_{2k_0+1}(t)=c_{2k_0+1}\frac{\tilde{m}(t)}{\tilde{m}(0)};\quad 
\chi_{2k_0+1}(t)=d_{2k_0+1}\frac{\tilde{q}(t)}{\tilde{q}(0)},
\end{eqnarray}
where $\tilde{m}(t)$ and $\tilde{q}(t)$ are nonzero solutions of the homogeneous equations \eqref{EQ_m} -- \eqref{EQ_q}. In this case, however, the boundary conditions \eqref{Inits_cd} on the coefficients of the \eqref{rho_chi_sol_series1} series change -- it is necessary to substitute zeros in the right parts of \eqref{Inits_cd} relations. In the case of the maximum degree of polynomials \eqref{rho_chi_sol_series1} $n_0=2k_0+1$, we obtain from \eqref{Inits_cd} a system of homogeneous linear algebraic equations with respect to the coefficients $\rho_{2k+1}(0)$\footnote{we consider only gravitational perturbations , since perturbations of a scalar field are no different in this sense}
\begin{eqnarray}\label{init_c,d}
\sum\limits_{k=0}^\infty \rho_{2k+1}(0)r_0^{2k+1}=0; \;\sum\limits_{k=1}^\infty k\rho_{2k+1}(0)r_0^{2k+1}=0.\nonumber\
\end{eqnarray}
Obviously, the matrix of this system has rank $\mathrm{ran}k=2$, so for the existence of a nontrivial solution it is necessary and sufficient that the number of coefficients be greater than two, i.e.,
\[k_0\geqslant 2,\]
and this, in turn, means that the minimum degree of the polynomial must be equal to 5. Therefore, consider the case of a polynomial of degree five in the radial variable:
\begin{equation}\label{rho5}
\rho(r,t)=\rho_1(t)r+\rho_3(t)r^3+\rho_5(t)r^5.\nonumber\
\end{equation}
In this case, the boundary conditions \eqref{bound_0} -- \eqref{bound_1} take the form:
\begin{eqnarray}\label{rho5=0}
\rho_1(0)r_0+\rho_3(0)r_0^3+\rho_5(0)r_0^5=0;\nonumber\\\
\label{drho5=0}
\rho_3(0)+2\rho_5(0)r_0^2=0\Rightarrow 
\rho_3(0)=-2\rho_5(0)r_0^2;\; \rho_1(0)=\rho_5(0)r_0^4.\nonumber\
\end{eqnarray}

Thus, there are no localized non-singular perturbations described by polynomials of the third degree in the radial variable. The minimum degree of the polynomial in the radial variable for localized non-singular perturbations is 5.

For nonsingular localized perturbations with $n\geqslant5$, it is necessary to solve the same recursive equations \eqref{rho_2m+1} -- \eqref{chi_2m+1}, on the right side of which, instead of the functions $m(t)$ and $q( t)$ you must use the functions $\tilde{m(}t)$ and $\tilde{q}(t)$. Thus, the problem has been formally solved.

\section{Numerical modeling of mass and scalar charge evolution}
Since the main goal of this work is to develop a theoretical and mathematical model of the process of evolution of spherical perturbations in a cosmological medium of degenerate scalarly charged fermions and methods for studying this model, in this article we will restrict ourselves to an example of numerical simulation of this process using the mathematical model described above, without going into analysis of numerical simulation results. These issues will be the subject of a separate article.

\subsection{Dimension of functions and parameters}
For numerical simulation, the dimension of physical quantities and the possibility of forming dimensionless complexes are important. Analyzing the Einstein equations, the scalar field equations and the definitions of various quantities, one can come to the following conclusion regarding the dimension of the functions and parameters included in them: \footnote{Square brackets indicate the dimensions of quantities in units of time, dimensionless quantities have\\ dimension 1.}
\begin{eqnarray}
[a]=[\xi]=[\Phi]=[\varphi]=[\nu]=[\delta_z]=[v_r]=[\psi]=[1];\nonumber
\end{eqnarray}\vspace{-16pt}
\begin{eqnarray}
[H]=[Z]=[m_0]=[e^2]=[\pi^2_c]=\bigl[t^{-1}\bigr];\nonumber
\end{eqnarray}\vspace{-16pt}
\begin{eqnarray}\label{dimen}
[\alpha]=[\Lambda]=\bigl[t^{-2}\bigr];\; [\rho_n]=[\chi_n]=[t^{-n+1}];
\end{eqnarray}\vspace{-16pt}
\begin{eqnarray}
 [r]=[t]\;,[\phi]=[\mu]=[\rho]=[\chi] =[m]=[q]=[d_z]=[t].\nonumber
\end{eqnarray}

Further, it must be remembered that the article uses the Planck system of units (see the footnote \ref{Plank_units}), therefore, to reduce the results to the usual systems of units, quantities that have the dimension $[t^{-n}]$ must be multiplied by the Planck mass to the same power -- $m_{pl}^n$ (or divide by the Planck time to the same power -- $t_{pl}^n$). The value of speed must be multiplied by the speed of light.

\subsection{Complete system of equations and initial conditions for numerical modeling of mass and charge evolution}
Thus, the solutions of the linear system of homogeneous differential equations $\mathbf{S_1 :}$ \eqref{EQ_m}, \eqref{EQ_q} and $\mathbf{S_2 :}$ \eqref{rho_1} and \eqref{chi_1} are determined by the basic functions of the background solution $H(t)$, $\Phi(t)$ and $Z(t)$ based on the numerical solution of the system of nonlinear equations $\mathbf{S_0:}$ \eqref{dot_xi-dot_varphi} -- \eqref {dZ/dt} with integral condition \eqref{Surf_Einst1_0}. From the point of view of numerical integration, it is more convenient to form a general system of differential equations and integrate it simultaneously, which significantly reduces time costs.

So, let's form the following system of differential equations:
\[\left\{\begin{array}{ll}
\mathbf{S_0}: & \{\eqref{dot_xi-dot_varphi} - \eqref{dZ/dt}, \eqref{Surf_Einst1_0}\}\\
\mathbf{S_1}: & \{\eqref{EQ_m}, \eqref{EQ_q}\}\\
\mathbf{S_2}: & \{\eqref{rho_1}, \eqref{chi_1}\}\\
\end{array}
\right.
\]
This system of equations has the following hierarchy $\mathbf{S_0}\to \mathbf{S_1}\to \mathbf{S_2}$: the system of background equations $\mathbf{S_0}$ is autonomous, the system of equations for singular mass and charge
$\mathbf{S_1}$ is determined by the solutions of $\mathbf{S_0}$, the system of equations for the functions $\rho_1(t)$ and $\chi_1(t)$, $\mathbf{S_2}$, is determined by the solutions of the systems $\ mathbf{S_0}$ and $\mathbf{S_1}$. In doing so, we will use the following initial conditions for the $\mathbf{S_0}$ subsystem, taking advantage of its autonomy (see \cite{Ignat_GC21})
\begin{equation}\label{IC_I}
\xi(0)=0;\; \Phi(0)=\Phi_0;\; Z(0)=0,
\end{equation}
 and the initial value of the Hubble parameter $H(0)$ will be defined as the positive root of the integral \eqref {Surf_Einst1_0} when initial conditions \eqref{IC_I} are substituted into it. Further, for simplicity, we assume that the initial values of the first derivatives of the functions $\mu(t),q(t),\rho_1(t),\chi_1(t)$ are zero.

When integrating the $\{\mathbf{S_1},\mathbf{S_2}\}$ system numerically, a natural question arises about the uniqueness of its solutions. As is known, the Cauchy problem for a system of ordinary linear differential equations of the second order always has a unique solution, provided that the coefficients of this system are continuous and bounded. In our case, the coefficients of the system are determined by the background solutions $\{\xi(t),H(t),\Phi(t),Z(t)\}$. The continuous dependence and boundedness of the coefficients of the $\mathbf{S_1}$ system depending on the values of the functions $\{\xi,H,Z\}\in \mathbb{Z}$ is obvious. The only danger may come from the background function of the potential $\Phi\to0$ due to the presence of this function in the denominator $\psi$ \eqref{psi}. However, a detailed study of the coefficients of the system shows that they are all continuous and finite as $\Phi\to0$. The continuous dependence of solutions on the parameters of the system requires a continuous dependence of the coefficients of these equations on the parameters. In this case, the danger may come from the charge parameter $e$. However, in this case too, by calculating the corresponding limits, one can prove the continuous dependence of the coefficients of the $\mathbf{S_1}$ system on all parameters.

Thus, we will define the system under study by the following ordered set of parameters and nontrivial initial conditions:
\begin{eqnarray}\label{params}
\mathbf{P}=[[\alpha,m_0,e,\pi_c],\Lambda];\\
\label{IC}
\mathbf{I}=[\Phi(0)=\Phi_0,m(0)=M_0,q(0)=Q_0,\rho_1(0)=\rho^0_1,\chi_1(0)=\chi^0_1,r_0].
\end{eqnarray}
\subsection{Model problem\label{mod2}}
First, let's solve two model problems, in which \emph{set} background functions:
\begin{eqnarray}\label{M_1}
\mathbf{M_1:}& \xi=0.001t,\; H=0.001, \Phi=1,\; Z=0\\
\label{M_2}
\mathbf{M_2:}& \xi=0.000001t,\; H=0.000001, \Phi=1,\; Z=0
\end{eqnarray}
and set, for simplicity, the cosmological constant equal to zero ($\Lambda=0$)
\begin{equation}\label{P_10}
\mathbf{P}_0=[[1,1,1,0.1],0].
\end{equation}
In doing so, we set the following initial conditions
\[M_0=1,\; Q_0=0,\; \left.\dot{m}\right|_{t=0}=0,\;\left.\dot{q}\right|_{t=0}=0, \]
corresponding to the initial singular perturbation mass $M_0=1\ m_{Pl}$ and to the zero initial perturbation singular scalar charge. Note that the model problem $\mathbf{M_1}$ corresponds to the value of the Hubble parameter $H=H_0=10^{-3}$, and the model problem $\mathbf{M_2}$ corresponds to the value of the Hubble parameter $H=H_0=10 ^{-6}$.

On Fig. \ref{ris2} -- \ref{ris2a} shows the results of numerical integration for these models. The gray bar shows the region of required mass values for supermassive black hole nuclei $M_{bhs}$ \eqref{M_nc}.
\TwoFigs{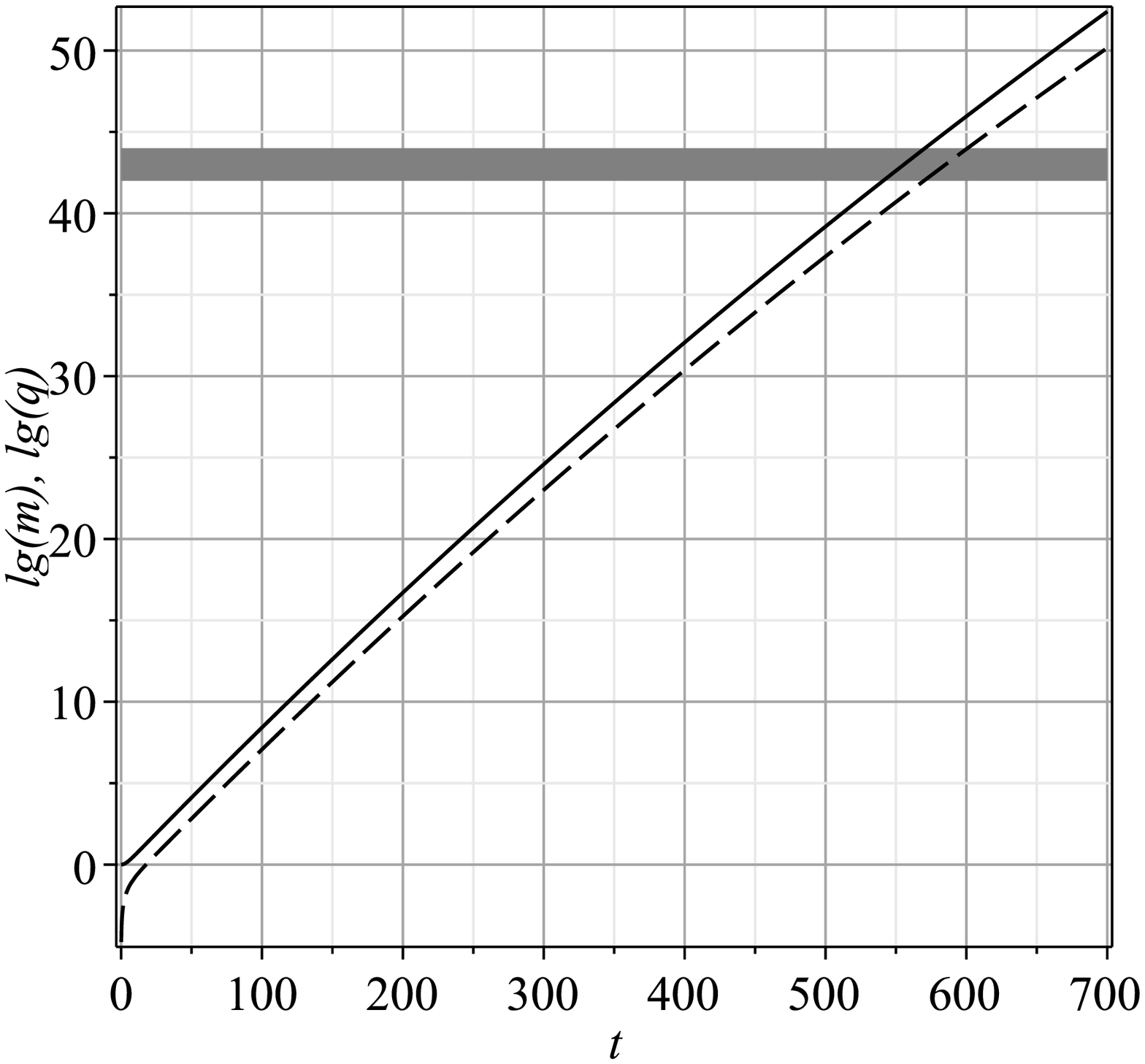}{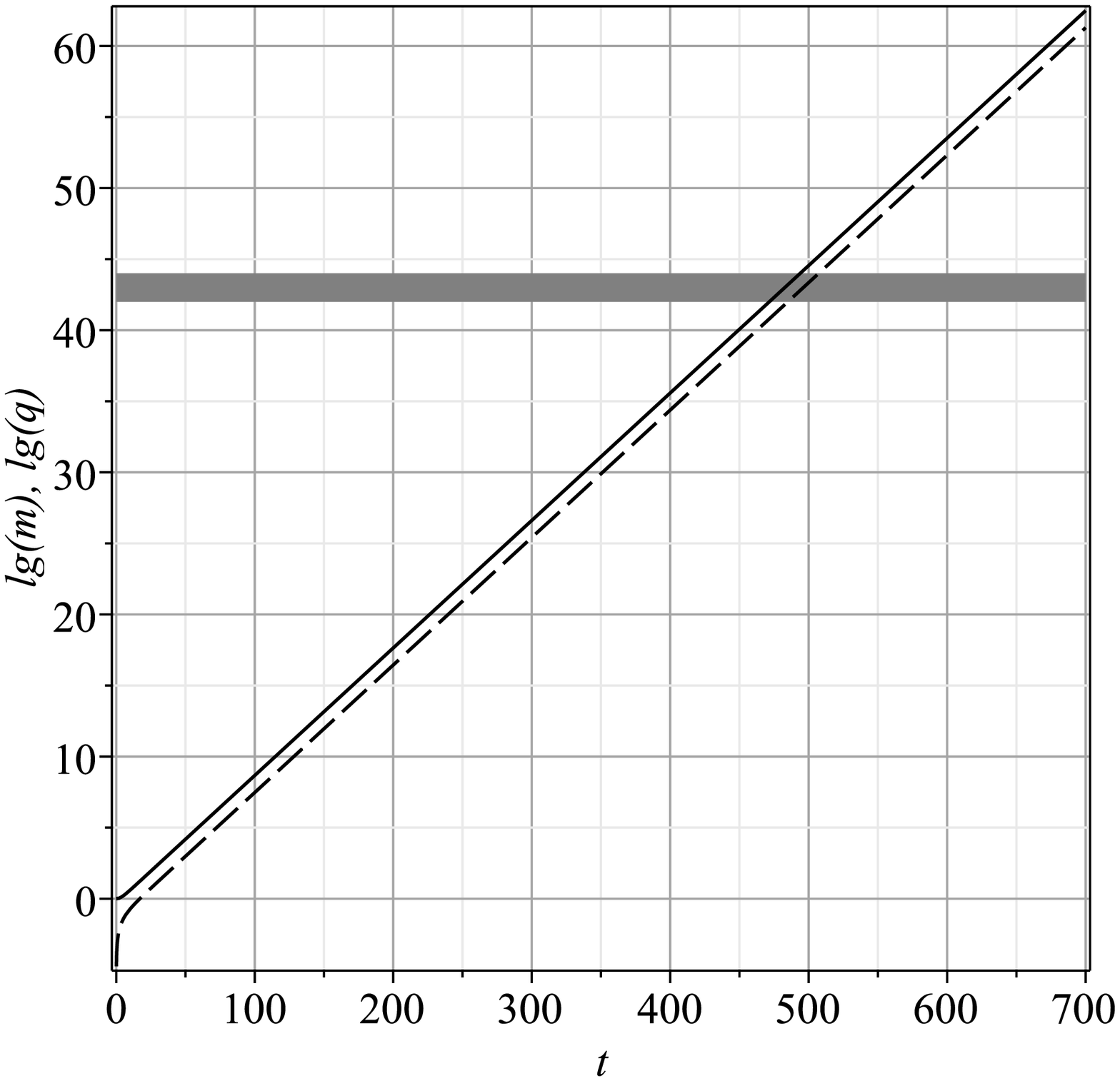}{\label{ris2} Evolution of the perturbation mass $m(t)$ (solid line) and charge $q(t)$ (dashed line) for the $\mathbf{M_1}$ \eqref{M_1}.}{\label{ris2a} model $ mass evolution m(t)$ perturbation (solid line) and charge $q(t)$ (dashed line) for the $\mathbf{M_2}$ \eqref{M_2} model.}
As can be seen from the graphs in Fig. \ref{ris2} -- \ref{ris2a}, the mass of the singular source $m(t)$, equal to $m_{pl}$ at the initial moment of time, grows exponentially rapidly over the entire interval and reaches a value of the order of $10^{52} \div 10^{62}$, the charge of the scalar source also grows exponentially rapidly, qualitatively repeating the behavior of the mass function. Moreover, in the $\mathbf{M_1}$ model with $H_0=10^{-3}$, the mass $M_{bhs}$ \eqref{M_nc} is reached at time $t_{bhs}\approx 550$, while in the $\mathbf{M_2}$ model with $H_0=10^{-6}$, the time to reach the mass $M_{bhs}$ is less -- $t_{bhs}\approx 470$. Note that in the $\mathbf{M_1}$ and $\mathbf{M_2}$ models, the number of e-folds required for inflationary cosmology is $N\gtrsim 60$ \cite{Rubak} (i.e., the value of the exponent argument $\ xi$) is reached at times $t_{in}\simeq 6\cdot 10^4 \div 6\cdot 10^7\ t_{Pl}$, which is much longer than the time $t_{bhs}$ (from 2 to 5 orders of magnitude ). This means that the process of formation of a supermassive black hole occurs at the stage of early inflation. These simple examples confirm our assumption about the possibility of an exponential growth of the mass of a spherical perturbation \cite{Yu_GC_3_22}, which is the motivation for integrating the full model.

On the other hand, the fact that the formation time of supermassive black holes in the early Universe is small compared to the standard inflation time $t_{bhs}\ll t_{in}$ indicates that the process of development of the scalar-gravitational instability in the early Universe is insensitive to the singularity of the cosmological model in general, i.e., to the type of gravity theory underlying the cosmological model. To maintain the normal speed of this process, only the very presence of inflationary expansion at the stage of development of instability is important.

Note that using the similarity property (Statement \ref{stat1}) we can propagate the result to other parameter values, for example:
\begin{equation}\label{P_11}
\mathbf{P}_{01}=[[10^{-8},10^{-4},1^{-2},10^{-3}],0]\Rightarrow \alpha=10^{-8}; m_0=10^{-4}; e=10^{-2}; \pi_c=10^{-3}.
\end{equation}
When extending the results of the subsection \textbf{\ref{mod2}} to the case \eqref{P_11}, one must also remember about the scaling transformation of coordinates, as a result of which the Hubble parameter must be transformed according to the law
\[H=\frac{d\ln a}{dt}\Rightarrow H\to kH.\]
Therefore, in the case of extending the results of the model problem with the parameters \eqref{P_10} to the case of \eqref{P_11}, we must also change the Hubble parameter in the $\mathbf{M_1}$ and $\mathbf{M_2}$ models to $10^{ -7}$ and $10^{-10}$, respectively. We also need to transform the time coordinate, in this case we get the characteristic value of the time of crossing the gray strip of the graphs in Fig. \ref{ris2} -- \ref{ris2a} $t_{bhs}\backsim 7\cdot10^{6}t_{Pl}$. Thus, the similarity property allows us to study the problem for parameter values of the order of unity, when this problem is accessible to standard numerical methods, and then to transfer the results obtained to the region of physically interesting small parameter values.
\subsection{Results of numerical simulation of mass and charge evolution}
In this article, we restrict ourselves to two illustrative examples of the numerical integration of the $\{\mathbf{S_0},\mathbf{S_1}\}$ system, leaving the numerical integration of the $\mathbf{S_2}$ system to the special article mentioned above. This is caused not so much by the technical problems of the process of numerical integration of a rather cumbersome system of differential equations, but by the detected strong dependence of the behavior of solutions on the system parameters and initial conditions. Note that the considered examples demonstrate only the main tendencies of the dependence of the process of formation of black holes on the fundamental parameters of the model, while not pretending to their physical significance.
\subsubsection{Small scalar charge}
Consider the case of the following parameter values \eqref{params} and initial conditions \eqref{IC} of the system under study:
\begin{eqnarray}\label{params1}
\mathbf{P_1}=[[1,1,0.0001,0.1],0];\\
\label{IC1}
\mathbf{I_1}=[\Phi(0)=1,m(0)=1,\nonumber\\
q(0)=0,\rho_1(0)=\rho^0_1,\chi_1(0)=\chi^0_1,r_0=10^4];\\
\label{IC2}
\mathbf{I_2}=[\Phi(0)=1,m(0)=0,\nonumber\\
q(0)=1,\rho_1(0)=\rho^0_1,\chi_1(0)=\chi^0_1,r_0=10^4]
\end{eqnarray}
Note that the solutions of the $\mathbf{S_0}$ background model do not depend on the initial conditions for perturbations, so the conditions \eqref{IC1} and \eqref{IC2} are equivalent with respect to the background solution. The order of the localization radius $r_0=10^{4}$ that we have chosen corresponds to the size $10^4 m^{-1}_{pl}\backsim 10^{-29}cm$ at time zero. Thus, the initial radius of perturbation localization is of the order of the size of Grand Unified particles at the energy level of the order of $10^{15}Gev$. Over time, the localization radius evolves according to the law $R=r_0 \exp(\xi(t))$. In addition, note that the initial conditions \eqref{IC1} and \eqref{IC2} correspond to the initial perturbation mass $m(0)=1$ and $q(0)=1$, i.e., the mass in $m (0)=m_{pl}$ and a similar charge.

\TwoFigs{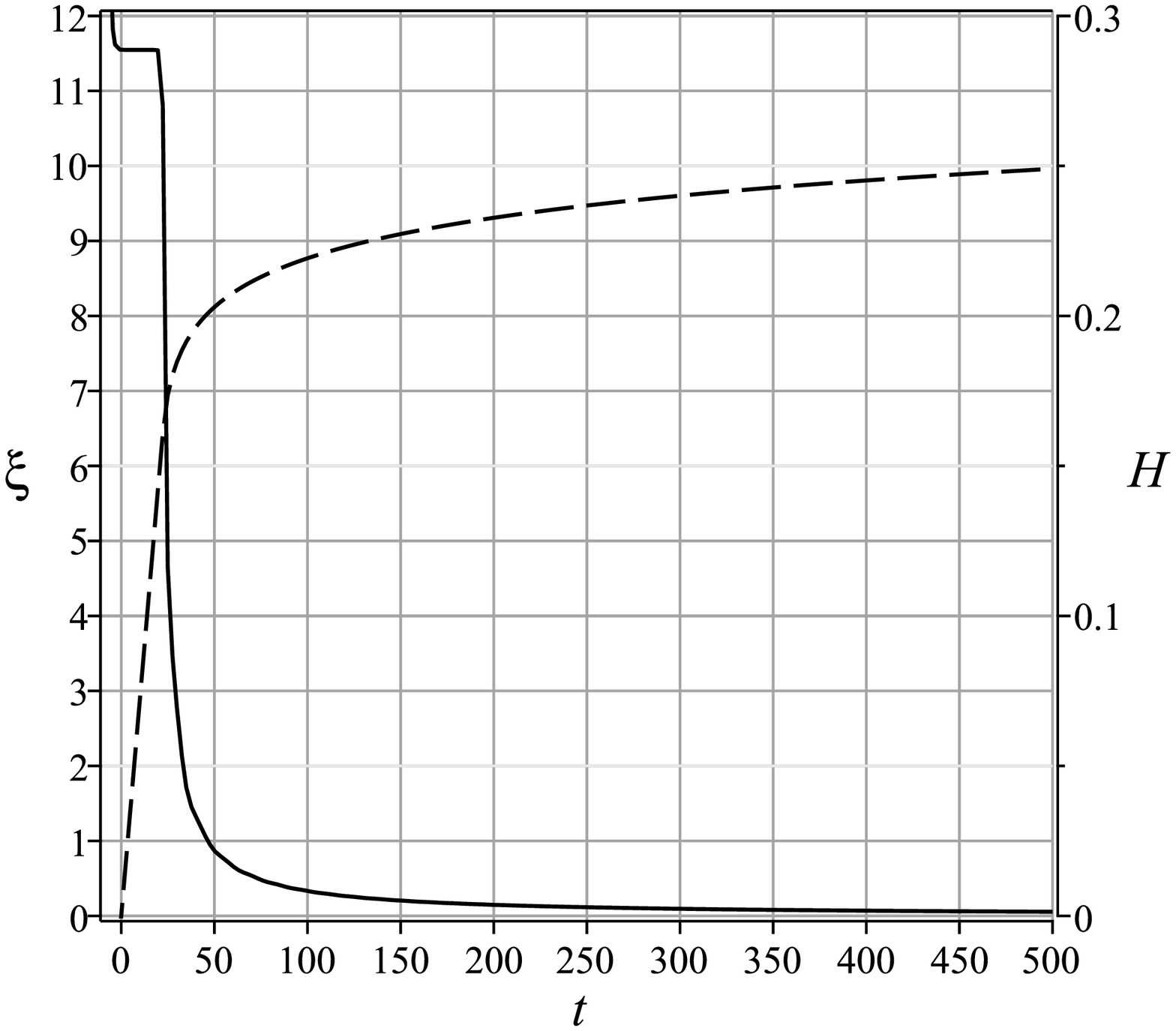}{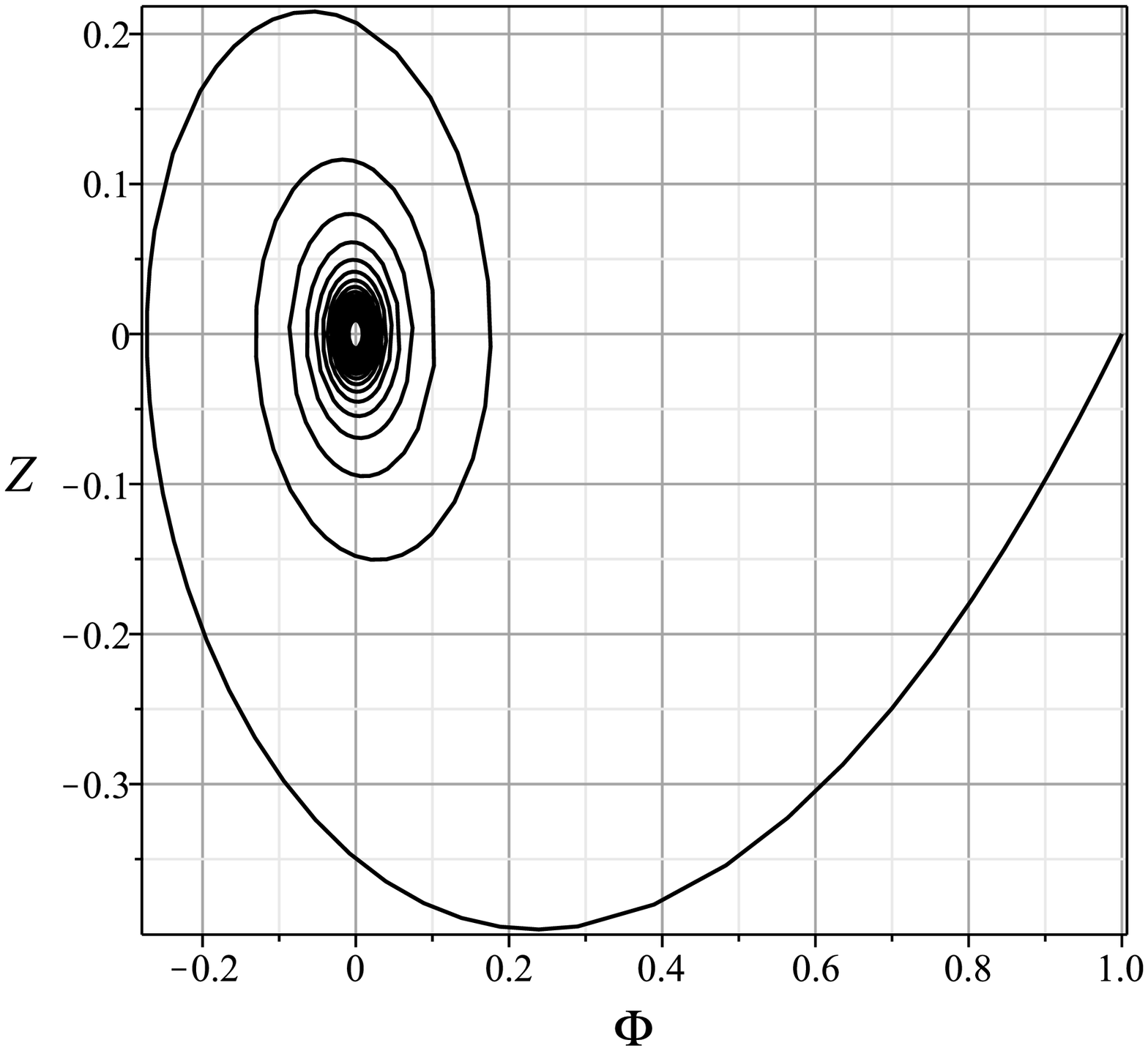}{\label{ris3} Evolution of the scale function $\xi(t)$ (dashed line) and the Hubble parameter $H(t)$ (solid line) for a model with \eqref{params1} parameters and initial conditions \eqref{IC1}.}{\label{ris4} Phase portrait of the model with parameters \eqref{params1} and initial conditions \eqref{IC1}.}

On Fig. \ref{ris3} and \ref{ris4} shows the evolution of the background functions of the unperturbed cosmological model corresponding to the values of the parameters \eqref{params1} and the initial conditions \eqref{IC1} with zero value of the cosmological constant. The background cosmological model starts from the time $t_0\approx -8.368$ in the time scale chosen by us, while the Hubble parameter after a short phase of early inflation rapidly falls and tends asymptotically to zero. At the same time, the phase trajectory of the scalar field winds asymptotically around the zero singular stable point of the dynamical system. Note that in models with charged fermions, the Higgs potential minimum points are not stable, the stable points are points with zero scalar potential \cite{TMF_21}, while the Higgs potential minimum points become saddle - they are only conditionally stable in one of the projections of the phase space. The stable points correspond to the nonrelativistic equation of state for fermions.

On Fig. \ref{ris5} and Fig. \ref{ris6} graphs of evolution of point mass $m(t)$ and point charge $q(t)$ are shown. As can be seen from the graphs presented, the value of the central mass $m(t)$ after a short-term surge corresponding to the stage of early inflation, first quickly falls into the region of negative values, and then slowly spreads. The magnitude of the central charge rapidly oscillates near zero.

\TwoFigs{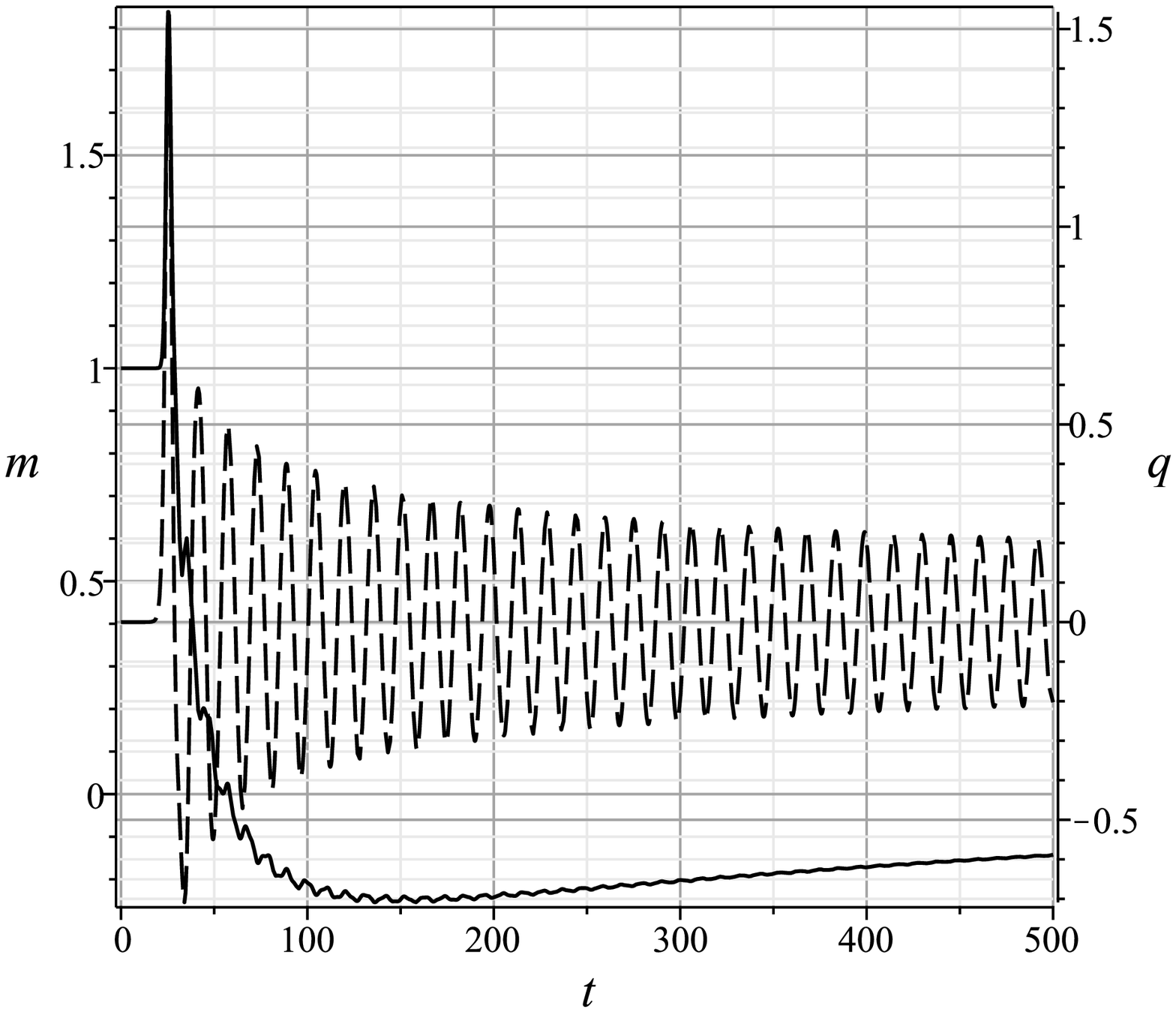}{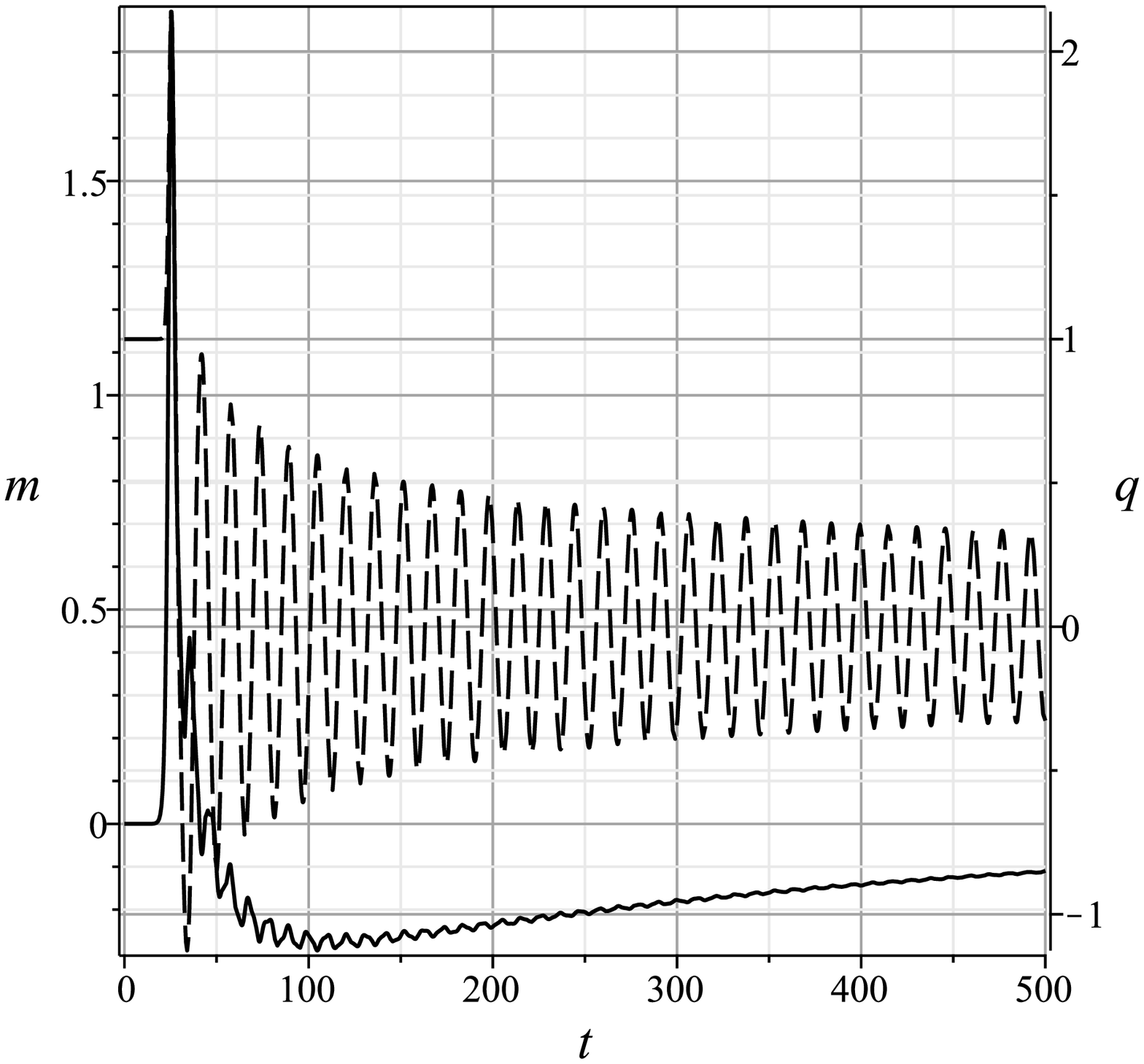}{\label{ris5} Evolution of perturbation mass $m(t)$ (solid line) and charge $q(t)$ (dashed line) for a model with parameters \eqref{params1} under initial conditions \eqref{IC1}}{\label{ris6} Evolution perturbation mass $m(t)$ (solid line) and charge $q(t)$ (dashed line) for a model with parameters \eqref{params1} under initial conditions \eqref{IC2}}

Note, firstly, that the point mass $m(t)$ corresponds only to a perturbation of the energy density, and therefore can also take negative values.

Secondly, note that the point charge $q(t)$ is not required to conserve, since it is generated by the scalar charge density $\sigma$ \eqref{2_3a_2}, which is not related to the scalar charge conservation law, in contrast to the kinematic density of the scalar charge $ \rho$ \eqref{2c}\footnote{See \cite{TMF_21} for details.}.

Third, let's pay attention to the fact that the graphs in Fig. \ref{ris5} correspond to zero initial conditions for a point charge -- $q(0)=0,\dot{q}(0)=0$, while the graphs in Fig. \ref{ris6} correspond to zero initial conditions for a point mass -- $m(0)=0,\dot{m}(0)=0$.

On these graphs one can observe the same qualitative behavior of the corresponding functions $m(t)$ and $q(t)$. Thus, on the one hand, we can conclude that a scalar charge is generated by a perturbation of mass and, conversely, that a perturbation of a mass is generated by a scalar charge, as we discussed above. On the other hand, taking into account the fast oscillatory nature of $m(t)$ and $q(t)$, we can conclude that the system quickly forgets its initial state, so the nature of the initial perturbation becomes secondary.

Thus, in the case of a cosmological model with small scalar charges (parameters \eqref{params1}), there is no significant growth of spherical scalar-gravitational perturbations.
\subsubsection{Large scalar charge}
In this regard, consider the case of the following values of the parameters \eqref{params} and initial conditions \eqref{IC} of the system under study, increasing the value of the scalar charge $e$ by 4 orders of magnitude:
\begin{eqnarray}\label{params2}
\mathbf{P_1}=[[1,1,1,0.1],0].
\end{eqnarray}
At the same time, we will keep the initial conditions \eqref{IC1}.

On Fig. \ref{ris7} and \ref{ris8} the evolution of the background functions of the unperturbed cosmological model corresponding to the values of the parameters \eqref{params2} and the initial conditions \eqref{IC1} with zero value of the cosmological constant is shown. The background cosmological model starts from the time $t_0\approx -6.719$ on the time scale chosen by us, while the Hubble parameter after a short phase of early inflation rapidly decreases and asymptotically tends to the value $H\approx 0.0197$. At the same time, the phase trajectory of the scalar field is asymptotically wound around the zero singular point (see, for example, \cite{YuKokh_TMF}).
\TwoFigs{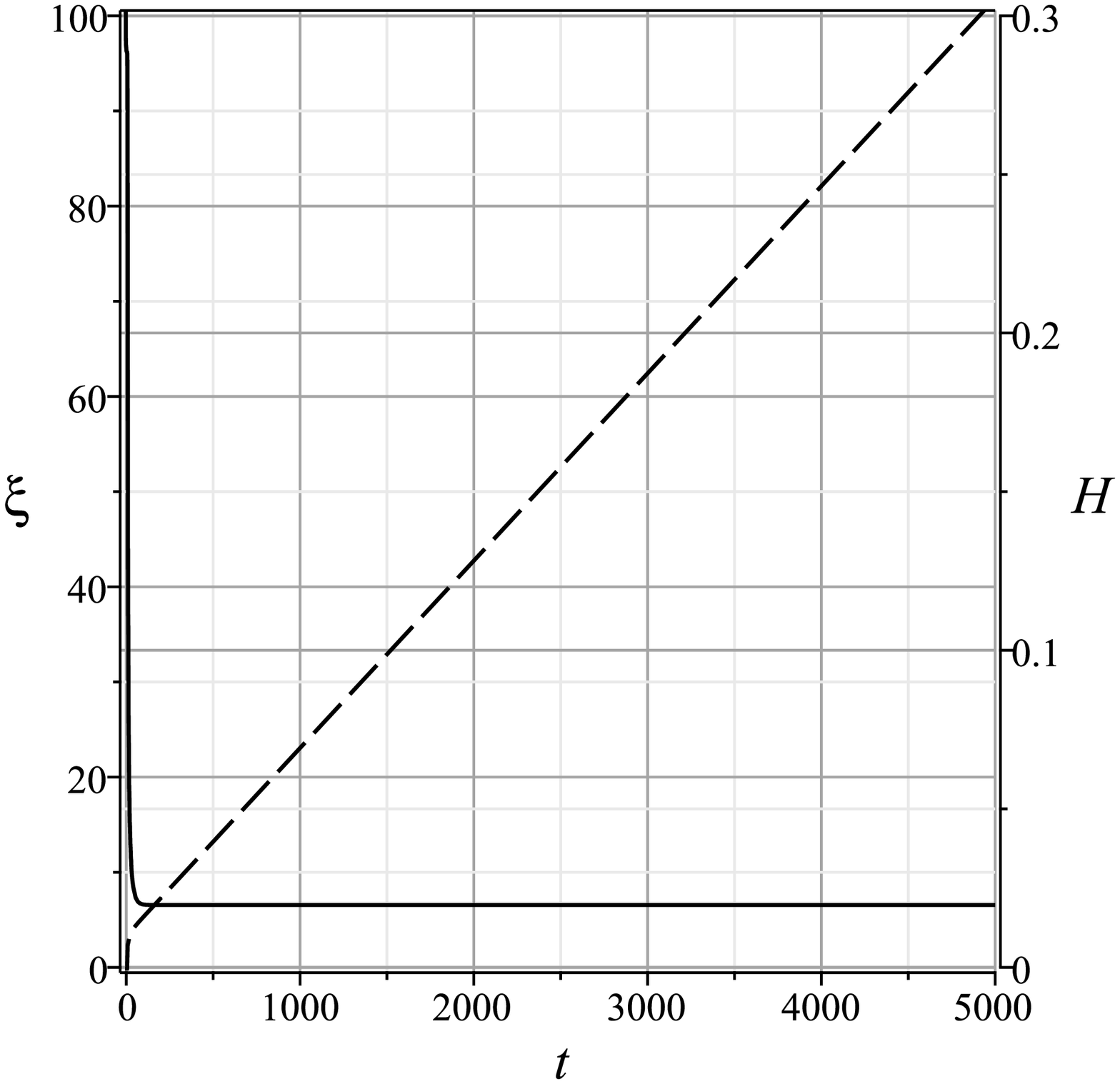}{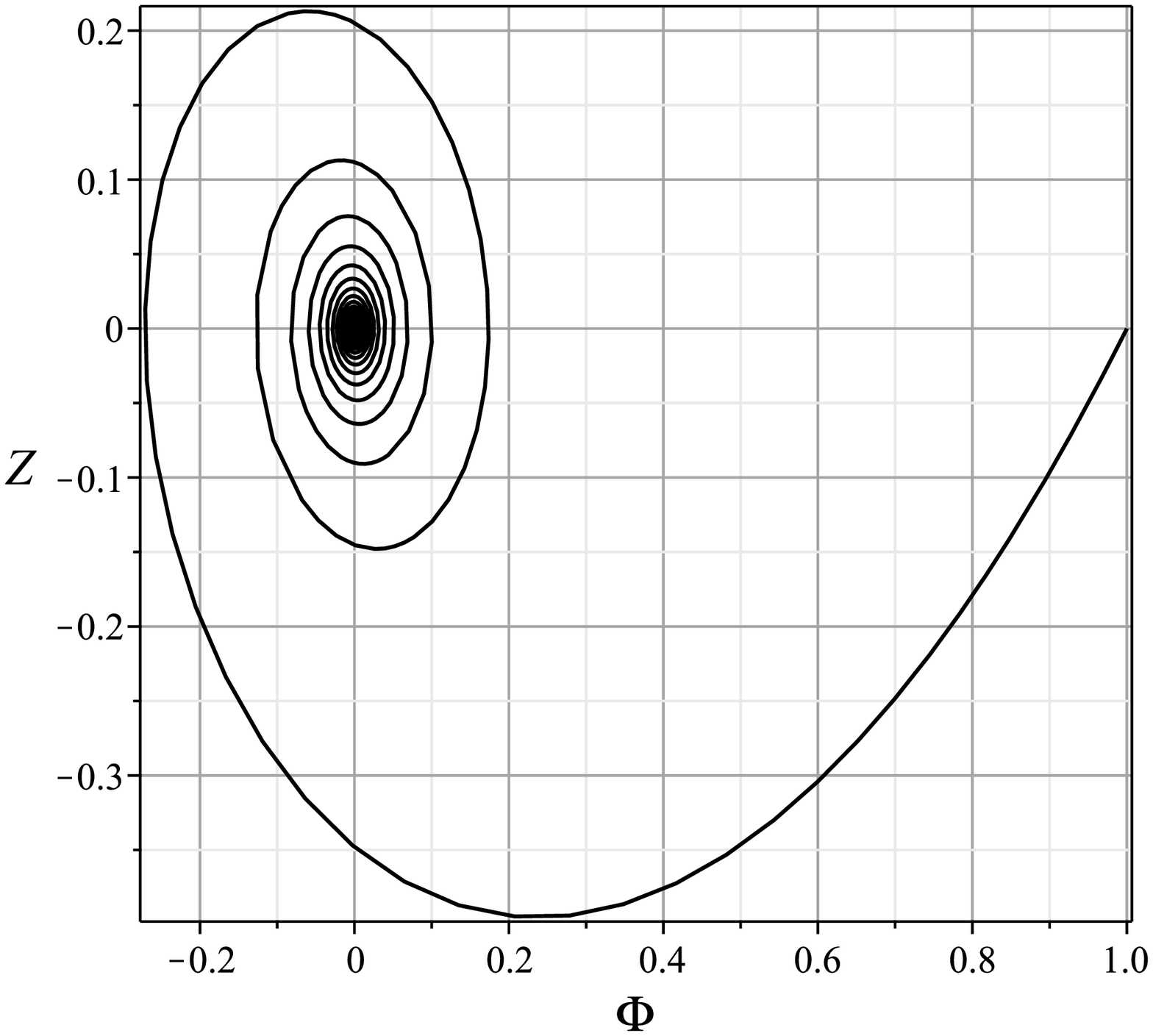}{\label{ris7} Evolution of the scale function $\xi(t)$ (dashed line) and the Hubble parameter $H(t)$ (solid line) for a model with \eqref{params2} parameters and initial conditions \eqref{IC1}.}{\label{ris8} Phase portrait of the model with parameters \eqref{params2} and initial conditions \eqref{IC1}.}

On Fig. \ref{ris9} the evolution of the perturbation mass and charge for this case is shown. As can be seen from the graphs, the singular mass $m(t)$ reaches the required values of the BHS mass \eqref{M_nc} already at $t\sim 230 t_{pl}$, starting from $M_0=1\ m_{pl}$ . The gray bar shows the region of required mass values for supermassive black hole nuclei $M_{bhs}$ \eqref{M_nc}.

For $t>100$, the function $\xi(t)$ grows linearly with time, which corresponds to inflation, while the number of e-folds required for inflationary cosmology is $N\gtrsim 60$ according to the graph in Fig. \ref{ris7} is reached at $t\backsimeq 3000$. This value is an order of magnitude greater than the above time required for the formation of a supermassive black hole. Therefore, it can be argued that in the considered model, the formation of supermassive black holes should occur at the early stages of inflation.

\fig{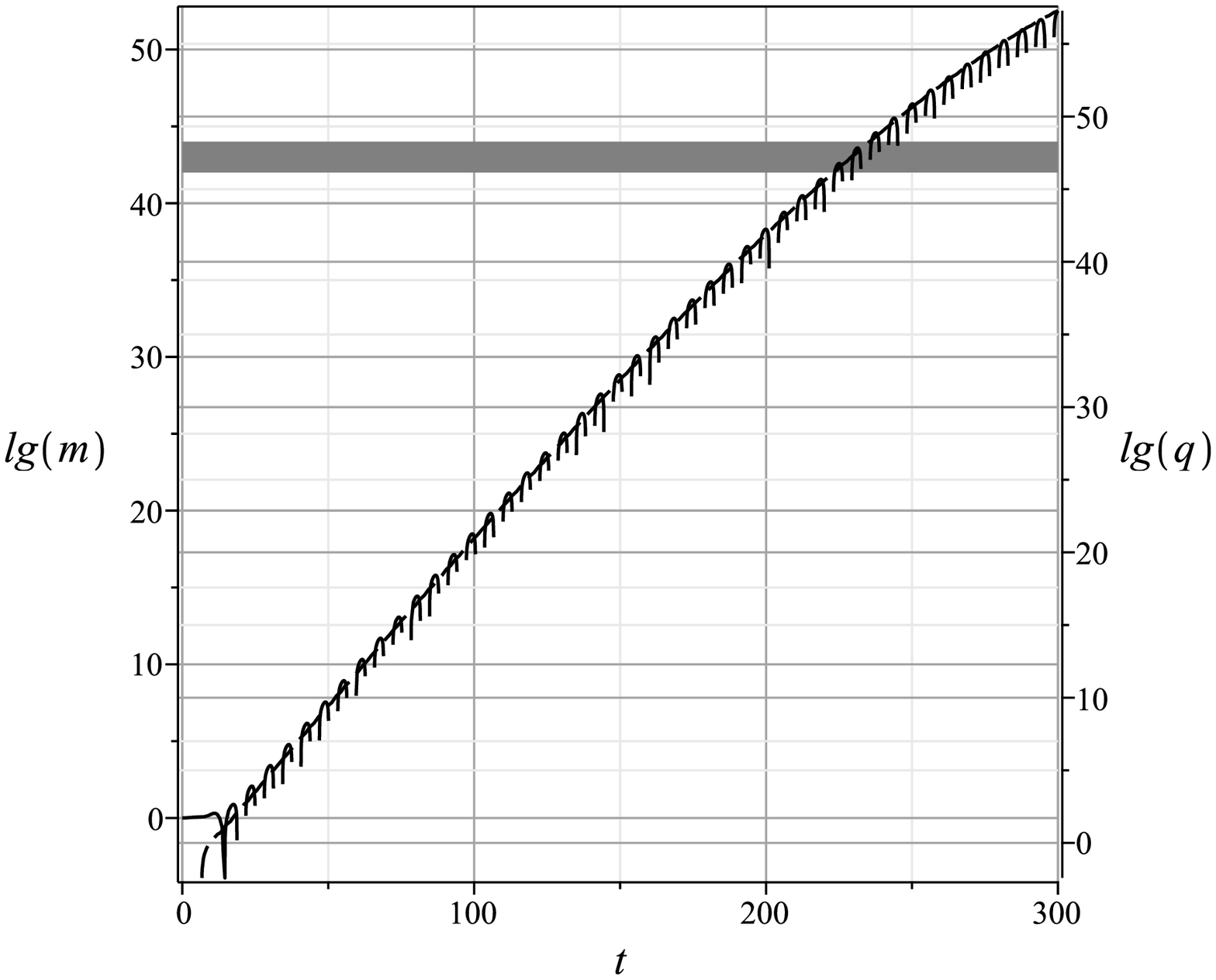}{10}{8}{\label{ris9}Evolution of perturbation mass $m(t)$ (solid line) and charge $q(t)$ (dashed line) for a model with parameters \eqref{params2} under initial conditions \eqref{IC1}}

\section{Conclusion}
In conclusion, we highlight some of the results obtained in this article.
%

%
Main results\\
On the basis of the theory of gravitating statistical systems with scalar interaction of particles, a mathematical model of the evolution of spherically symmetric perturbations in cosmological matter, consisting of degenerate scalar charged fermions with Higgs scalar interaction, is constructed and studied.\\
\stroka{The mathematical model consists of three subsystems of system of differential equations, in which the first autonomous subsystem $\mathbf{S_0}$ \ describes the unperturbed cosmological model and, in turn, consists of four nonlinear ordinary differential equations \eqref{dot_xi-dot_varphi} -- \ eqref{dZ/dt} and their first normalized integral \eqref{Surf_Einst1_0}, the second subsystem describes the evolution of the total mass and charge corresponding to the singular part of spherical perturbations, and represents a system of two linear homogeneous ordinary differential equations of the second order $\mathbf{S_1 :} $ \eqref{EQ_m}, \eqref{EQ_q} and, finally, the third subsystem describes the evolution of nonsingular parts of perturbations and represents a system of two homogeneous linear partial differential equations \eqref{EQ_rho} -- \eqref{EQ_chi}. The coefficients of all equations depend on time and are determined by the background solutions of the subsystem $\mathbf{S_0}:$ \eqref{dot_xi-dot_varphi} -- \eqref{dZ/dt}.}
\stroka{On the basis of the obtained equations, the problem of the evolution of perturbations spatially localized at the initial moment of time, vanishing together with the first derivatives with respect to the radial variable on a sphere of a given radius, is solved. The solution can be represented by polynomials in a radial variable with odd degrees. In this case, the polynomial coefficients represent time functions that satisfy the system of recurrent nonhomogeneous linear ordinary differential equations \eqref{rho_2m+1} -- \eqref{chi_2m+1}.}
\stroka{In a particular, physically significant case of a cubic polynomial, the system of equations for the non-singular part of perturbations is reduced to a system of two inhomogeneous second-order linear differential equations, on the basis of which the preservation of the coordinate radius of perturbation localization in the process of evolution is shown. In this case, the physical radius of localization evolves in proportion to the scale factor.}
\stroka{The numerical integration of the first two subsystems of the model in the case of a polynomial perturbation of the third degree is carried out, on the basis of which the exponential growth of the central mass of the perturbation and the oscillatory nature of the growing charge are demonstrated.}
\stroka{Thus, the study carried out in the article, firstly, confirmed the presence of strong instability and spherically symmetric perturbations, and, secondly, gave a closed theoretical model for a detailed study of the process of early formation of supermassive black holes. We plan to carry out such studies in future works.
}

In conclusion, we make three remarks both about the generality of the methods developed in it and about the possible development of its results.\\
\hskip 12pt First, we note that the methods of separating variables and extracting the singular part of perturbations developed in Sections 2 and 3 can also be extended to other cosmological models based on alternative gravity models, if these models are described by second-order differential equations. Of course, provided that these theories admit astrophysical spherically symmetric solutions, in particular, they allow black hole-type solutions. Since the existence of black holes has been confirmed by numerous observations over the past 6 years, it seems that the requirement for the existence of such solutions should be a sine qua non for any viable theoretical model of gravity. Further, the equations of the second order \emph{linear perturbation theory} in the case of spherical symmetry must contain the spherical Laplass operator. Therefore, the evolution equations for linear spherical perturbations in the correct gravity model \emph{in their structure} cannot differ from the evolution equations \eqref{Einst_dp} -- \eqref{Eq_varphi}. In these equations, only the time-dependent coefficients can change at the first and zero derivatives of perturbations. Thus, the applicability of the method of separation of variables and selection of the singular part of perturbations, which is directly related to the Schwarzschild solution, i.e., to the existence of black holes, will also remain.\\
\hskip 12pt Secondly, as is known, one of the main reasons for the appearance of modern modifications of the theory of gravity based on second-order equations, for example, scalar-tensor theories, theories of gravity with a non-minimal theory of gravity, at the early and late stages of the evolution of the universe. Due to the selection of the constants of the theory, in the Horndesky model, in particular, it is possible to solve this problem. However, we must understand that the conclusions of these theories begin to diverge from the conclusions of Einstein's theory only at very large cosmological times, corresponding to the current era of the evolution of the Universe. The high rate of development of the scalar-gravitational instability compared to the cosmological rates in the early Universe makes it possible to ignore the features of gravity models when studying the formation of supermassive black holes in the early Universe.\\
\hskip 12pt Thirdly, we note that if the early formation of SBH (see page \pageref{_SBH}) is an inevitable stage in the evolution of the Universe, then this process can radically change the nature of its further evolution. Indeed, according to the results of this article, after the SBH formation process, the scalar field can be localized inside these black holes. In this case, the Universe after the stage of formation of SBH can be a mixture of a heated relativistic gas of particles with inclusions of SBH in it. The scalar field will disappear and the Universe will evolve with the Hubble parameter determined by the quadratic gravitational fluctuations introduced by SBH \cite{Yu_BH_19} and the effective cosmological constant
\[\Lambda_1=\frac{9n}{4r^2_0}\left(\frac{2\mu}{r_0}\right)^2,\]
where $n$ is the density of the number SBH, $\mu$ and $r_0$ are their reduced masses and radii.

\section*{Thanks}
The author is grateful to the participants of the seminar of the Department of Theory of Relativity and Gravity of Kazan University for the useful discussion of the work. The author is especially grateful to professors S.V. Sushkov and A.B. Balakin for very useful comments and discussion of the features of modern modifications of the theory of gravity as applied to cosmology and astrophysics. The Author is also grateful to the participants of the 5th International Winter School-Seminar <<Peter's Readings-2022>> for the fruitful discussion of the report (11/24/22), and especially to Professors K.A. Bronnikov, M.O. Katanaev and B. Sakha, as well as associate professor P.I. Pronin.

\section*{Founding}
The work was carried out at the expense of a subsidy allocated as part of the state support of the Kazan (Volga Region) Federal University in order to increase its competitiveness among the world's leading scientific and educational centers.

\end{document}